\documentclass[longauth]{aa} % for the long lists of affiliations 
\usepackage{color}
\usepackage{graphicx}
\usepackage{txfonts}

\newcommand{\ms}{m\,s$^{-1}$}

\def\vsini{\ensuremath{{\upsilon}\sin i}}
\def\kms{$\mathrm{km\,s}^{-1}$}
\def\ms{\hbox{\,m\,s$^{-1}$}}         %m.s -1
\def\m2s2{\hbox{\,m$^{2}$\,s$^{-2}$}} %m2.s -2
\def\kms{\hbox{\,km\,s$^{-1}$}}       %km.s -1
\def\vsini{\hbox{$v$\,sin\,$i_s$}}      %vsini
\def\rs{R$_{\rm s}$}

\begin{document} 

   \title{The GAPS Programme with HARPS-N at TNG}
   \subtitle{XXXI. The WASP-33 system revisited with HARPS-N\thanks{Based on observations made with the Italian Telescopio Nazionale Galileo (TNG) operated on the island of La Palma by the Fundacion Galileo Galilei of the INAF at the Spanish Observatorio Roque de los Muchachos of the IAC in the frame of the program Global Architecture of the Planetary Systems (GAPS).}
}
   \titlerunning{The WASP-33 system revisited with HARPS-N}
   \authorrunning{F.~Borsa et al.}
   \author{F.~Borsa\inst{\ref{brera}}, 
         A.~F.~Lanza\inst{\ref{catania}},
       I.~Raspantini\inst{\ref{milanouniv}},
               M.~Rainer\inst{\ref{arcetri}},  
              L.~Fossati\inst{\ref{graz}}, 
        M.~Brogi\inst{\ref{warwick},\ref{torino},\ref{warwick2}}, 
      M.~P.~Di Mauro\inst{\ref{iaps}},
      R.~Gratton\inst{\ref{padova}},
      L.~Pino\inst{\ref{arcetri}},
      S.~Benatti\inst{\ref{palermo}}, A.~Bignamini\inst{\ref{trieste}},  A.~S.~Bonomo\inst{\ref{torino}},  R.~Claudi\inst{\ref{padova}}, M.~Esposito\inst{\ref{tautenburg}}, 
      G.~Frustagli\inst{\ref{milanouniv},\ref{brera}},  A.~Maggio\inst{\ref{palermo}},  J.~Maldonado\inst{\ref{palermo}}, L.~Mancini\inst{\ref{romauniv},\ref{heidelberg},\ref{torino}},  G.~Micela\inst{\ref{palermo}}, 
      V.~Nascimbeni\inst{\ref{padova}},  E.~Poretti\inst{\ref{brera},\ref{tng}}, G.~Scandariato\inst{\ref{catania}},  D.~Sicilia\inst{\ref{catania}}, A.~Sozzetti\inst{\ref{torino}}, 
      W.~Boschin\inst{\ref{tng},\ref{IAC},\ref{IAC2}}, R.~Cosentino\inst{\ref{tng}}, E.~Covino\inst{\ref{napoli}}, S.~Desidera\inst{\ref{padova}}, L.~Di~Fabrizio\inst{\ref{tng}},  A.~F.~M.~Fiorenzano\inst{\ref{tng}},   A.~Harutyunyan\inst{\ref{tng}}, C.~Knapic\inst{\ref{trieste}}, E.~Molinari\inst{\ref{cagliari}}, I.~Pagano\inst{\ref{catania}}, M.~Pedani\inst{\ref{tng}}, G.~Piotto\inst{\ref{padovauniv}}     
          }
           
   \institute{INAF -- Osservatorio Astronomico di Brera, Via E. Bianchi 46, 23807 Merate (LC), Italy \label{brera}
   \and
   INAF -- Osservatorio Astrofisico di Catania, Via S.Sofia 78, 95123, Catania, Italy  \label{catania}
   \and
Dipartimento di Fisica, Universit\`{a} degli Studi di Milano Bicocca, Piazza dell'Ateneo Nuovo, 1, I-20126 Milano, Italy \label{milanouniv}
\and
INAF -- Osservatorio Astrofisico di Arcetri, Largo E. Fermi 5, 50125 Firenze, Italy \label{arcetri}
\and
Space Research Institute, Austrian Academy of Sciences, Schmiedlstrasse 6, A-8042 Graz, Austria  \label{graz}
\and
Department of Physics, University of Warwick, Coventry CV4 7AL, UK  \label{warwick}
\and
INAF -- Osservatorio Astrofisico di Torino, Via Osservatorio 20, 10025, Pino Torinese, Italy  \label{torino}
\and
Centre for Exoplanets and Habitability, University of Warwick, Gibbet Hill Road, Coventry CV4 7AL, UK  \label{warwick2}
\and
INAF-IAPS Istituto di Astrofisica e Planetologia Spaziali, Via del Fosso del Cavaliere 100, 00133, Roma, Italy \label{iaps}
\and
INAF -- Osservatorio Astronomico di Padova, Vicolo dell'Osservatorio 5, 35122, Padova, Italy  \label{padova}
\and
INAF -- Osservatorio Astronomico di Palermo, Piazza del Parlamento, 1, 90134, Palermo, Italy  \label{palermo}
\and
INAF -- Osservatorio Astronomico di Trieste, via Tiepolo 11, 34143 Trieste, Italy  \label{trieste}
\and 
Thüringer Landessternwarte Tautenburg, Sternwarte 5, 07778, Tautenburg, Germany  \label{tautenburg}
\and
Department of Physics, University of Rome Tor Vergata, Via della Ricerca Scientifica 1, I-00133 Rome, Italy  \label{romauniv}
\and
Max Planck Institute for Astronomy, K\"{o}nigstuhl 17, D-69117, Heidelberg, Germany  \label{heidelberg}
\and
Fundaci{\'o}n Galileo Galilei - INAF, Rambla Jos{\'e} Ana Fernandez P{\'e}rez 7, 38712 Bre$\tilde{\rm n}$a Baja, TF - Spain  \label{tng}
\and
Instituto de Astrof\'{\i}sica de Canarias (IAC), C/V\'{\i}a L\'actea s/n, 38205 La Laguna, TF - Spain \label{IAC} 
\and 
Departamento de Astrof\'{\i}sica, Universidad de La Laguna (ULL), 38206 La Laguna, TF - Spain \label{IAC2} 
\and
INAF -- Osservatorio Astronomico di Capodimonte, Salita Moiariello 16, 80131, Napoli, Italy  \label{napoli}
\and 
INAF -- Osservatorio di Cagliari, via della Scienza 5, I-09047 Selargius, CA, Italy  \label{cagliari}
\and
Dip. di Fisica e Astronomia Galileo Galilei -- Universit$\grave{\rm a}$ di Padova, Vicolo dell'Osservatorio 2, 35122, Padova, Italy  \label{padovauniv}
             }
             \offprints{F.~Borsa\\ \email{francesco.borsa@inaf.it}}

   \date{Received ; accepted }

% \abstract{}{}{}{}{} 
% 5 {} token are mandatory
 
  \abstract
   {Giant planets in short-period orbits around bright stars represent optimal candidates for atmospheric and dynamical studies of exoplanetary systems.}
   {We analyse four transits of WASP-33b observed with the optical high-resolution HARPS-N spectrograph to confirm its nodal precession, study its atmosphere and investigate the presence of star-planet interactions.}
   {We extract the mean line profiles of the spectra by using the Least Square Deconvolution method, and analyse the Doppler shadow and the radial velocities. 
   We also derive the transmission spectrum of the planet, correcting it for the stellar contamination due to rotation, center-to-limb variations and pulsations.}
   {We confirm the previously discovered nodal precession of WASP-33b, almost doubling the time coverage of the inclination and projected spin-orbit angle variation. 
   We find that the projected obliquity reached a minimum in 2011 and use this constraint to derive the geometry of the system, in particular its obliquity at that epoch ($\epsilon = 113.99^{\circ} \pm \, 0.22^{\circ}$) 
   and the inclination of the stellar spin axis ($i_{\rm s} = 90.11^{\circ}\, \pm \, 0.12^{\circ}$), as well as the gravitational quadrupole moment of the star $J_{2} = (6.73 \pm 0.22) \times 10^{-5}$, that we find to be in close agreement with the theoretically predicted value. 
   Small systematics errors are  computed by shifting the date of the minimum projected obliquity.
   We present detections of H$\alpha$ and H$\beta$ absorption in the atmosphere of the planet, with a contrast almost twice smaller than previously detected in the literature.
   We also find evidence for the presence of a pre-transit signal, which repeats in all four analysed transits and should thus be related to the planet. 
   The most likely explanation lies in a possible excitation of a stellar pulsation mode by the presence of the planetary companion.
     }
  % conclusions heading (optional)
   {A future common analysis of all available datasets in the literature will help shedding light on the possibility that the observed Balmer lines transit depth variations are related to stellar activity and/or pulsation, and to set constraints on the planetary temperature-pressure structure and thus on the energetics possibly driving atmospheric escape. A complete orbital phase coverage of WASP-33b with high-resolution spectroscopic (and spectro-polarimetric) observations could help understanding the nature of the pre-transit signal.
   }


   \maketitle
%
%________________________________________________________________

\section{Introduction\label{sec:intro}}
Transiting exoplanets orbiting intermediate mass (e.g., A-type) stars on short period ($P_{\rm orb}\leq$5 days) orbits are excellent laboratories for atmospheric and dynamical studies. 
The high dayside equilibrium temperatures of these Ultra-Hot Jupiters (UHJs, $T_{\rm eq}\geq$2200 K) cause the thermal dissociation of most molecules \citep[e.g.,][]{2018ApJ...866...27L,2018ApJ...855L..30A}. Depending on the heat transfer efficiency in the atmosphere there might be also variations in the chemical composition of the day-side, night-side and terminator of the planet \citep[e.g.,][]{2018A&A...617A.110P}.
Orbiting at short distances from their parent star, the strong stellar ultraviolet irradiation makes the planetary atmosphere undergoing a significant mass loss, affecting its composition
and evolution \citep[e.g.,][]{2018ApJ...868L..30F}.

Short-period systems are interesting also because they experience strong tidal interactions between the planet and its host star.
Star-planet tidal interactions could modify the stellar rotation rate along the system’s evolution \citep{2018A&A...619A..80G}, 
cause planet oblateness \citep{2019A&A...621A.117A}, suppress stellar activity \citep{2018AJ....155..113F}, and/or induce stellar pulsations \citep{2017ApJ...836L..17D}.
The measure of the projected spin-orbit angle either with radial velocities (RVs) through the Rossiter-McLaughlin effect \citep[RM,][]{1924ApJ....60...15R,1924ApJ....60...22M} or with the Doppler tomography technique \citep[e.g.,][]{2010MNRAS.403..151C} can help to shed light on a system evolution history \citep[e.g.,][]{2009ApJ...696.1230F}.

In this context, WASP-33b \citep{cameron} is a very intriguing target. It is a UHJ ($M_{\rm p}\sim$2.2 $M_{\rm Jup}$, $R_{\rm p}\sim$1.6 $R_{\rm Jup}$)
orbiting in $\sim$1.21 days a bright $\delta$-Scuti A-type star (V=8.3, $T_{\rm eff}\sim$7500 K, \vsini$\sim$86 \kms) with a highly misaligned projected spin-orbit angle $\lambda\sim$-110 degrees \citep[][]{cameron,2015A&A...578L...4L}.
Stellar non-radial pulsations were already noted spectroscopically in the discovery paper \citep{cameron}.
Photometric oscillations of the star were first reported by \citet{2011A&A...526L..10H}, and the stellar pulsation spectrum has been analysed in different studies \citep{Kovacsetal13,2014A&A...561A..48V,2015IAUGA..2255391M,2020arXiv200410767V}.
Because of its easily detected stellar pulsations, the WASP-33 system has been considered a good candidate for the detection of star-planet interactions \citep{2011A&A...526L..10H}.
WASP-33b was also noted as a possible target for which both classical and relativistic node precessional effects could be evidenced within a reasonable amount of time \citep{Iorio11,2016MNRAS.455..207I}; indeed classical node precessional effects have been afterwards detected by \citet{Johnsonetal15} and \citet{Watanabeetal20}. 

Its atmosphere has also been the subject of different studies.
The probable presence of a temperature inversion in its atmosphere \citep{2015ApJ...806..146H,2015A&A...584A..75V} was best explained by the presence of titanium oxide (TiO), whose detection with high-resolution spectroscopy is however debated \citep{2017AJ....154..221N,2020arXiv200610743H}.
Other results include the first indication of aluminum oxide (AlO) in an exoplanet by using low resolution spectrophotometry \citep{2019A&A...622A..71V}, 
the detection of ionized calcium (\ion{Ca}{ii} H\&K) up to very high upper-atmosphere layers close to the planetary Roche lobe \citep{2019A&A...632A..69Y} and of Balmer lines \citep{2020arXiv201107888Y,cauley2020}.
The existence of a thermal inversion was recently confirmed with the detection of \ion{Fe}{i} in emission using high-resolution observations \citep{2020ApJ...898L..31N}. 

Driven by the intriguing characteristics of the system, in this work we analyse new transits of WASP-33b taken with the HARPS-N high-resolution spectrograph, looking for confirmation of the nodal precession and exploring its atmosphere.
This manuscript is organized as follows. We first present our data in Sect.~\ref{sec:data_sample}, then analyse the planetary Doppler shadow and the in-transit RVs studying also the precession of the orbital plane and of the spin of the host star (Sect.~\ref{sec:mlp} and \ref{sec:rv}).
We study the planetary atmosphere focusing on H$\alpha$ and H$\beta$ absorption in Sect.~\ref{sec:Halpha_main}, finding a pre-transit signal coherent with the planetary orbital period discussed in Sect.~\ref{sec:pretransit}. 
Then we look for other atmospheric species with the cross-correlation technique in Sect.~\ref{sec:crosscorr}, ending with summary and conclusions in Sect.~\ref{sec:discussion}.

%__________________________________________________________________

\section{Data sample\label{sec:data_sample}}

We observed WASP-33 during four transits using the high-resolution (resolving power R$\sim$115000) HARPS-N spectrograph \citep{2012SPIE.8446E..1VC}, mounted at the Telescopio Nazionale Galileo on the La Palma island and covering the wavelength range 380-690 nm.
The first two transits were observed with HARPS-N only (program A34TAC42, PI Nascimbeni), 
while the other two were taken in the framework of the GAPS project \citep{2013A&A...554A..28C}. The latter were observed using the GIARPS configuration \citep{2017EPJP..132..364C}, but in this manuscript we focus only on the HARPS-N data.

The transit of WASP-33b lasts 2 hours 48 minutes. 
Three transit observations were monitored with exposures of 600 sec, while for one we used a longer cadence (900 sec exposures).
While the fiber A of the spectrograph was centered on the target, for all the transits the fiber B was monitoring simultaneously the sky to check for possible atmospheric emission contamination.
The log of the observations is reported in Table~\ref{tab:log}.

\begin{table}
\begin{center}
\caption{WASP-33 HARPS-N observations log.}
\label{tab:log}
\footnotesize
\begin{tabular}{ccccc}
 \hline\hline
 \noalign{\smallskip}
Transit number & Night$^{1}$ & Exposure time & N$_{\rm obs}$ & S/N$_{\rm ave}$\\
 \noalign{\smallskip}
 \hline
\noalign{\smallskip}
1 & 28 Sep 2016  &  600s & 40 & 107\\
2$^{*}$ & 20 Oct 2016 & 600s & 32 & 90\\
3 & 12 Jan 2018 & 900s & 23 & 166\\
4 & 02 Jan 2019 & 600s & 33 & 115\\
\noalign{\smallskip}
 \hline
  \multicolumn{3}{l}{$^{*}$\footnotesize{Weather conditions rapidly worsening.}} \\
    \multicolumn{3}{l}{$^{1}$\footnotesize{Start of night civil date.}} \\
\end{tabular}
\end{center}
\end{table}

The quality of transit 2 rapidly decreases after the transit ingress due to weather conditions, we thus decided to not include it in the analysis of the transmission spectrum, Doppler shadow and RVs, but only in Sect. \ref{sec:pretransit} where we look at the pre-transit portion of the data.

%______________________________________________________________
\section{Precession of the orbital plane and stellar spin axis\label{sec:mlp}}

\begin{figure*}%[!ht]
\centering
\includegraphics[width=\linewidth]{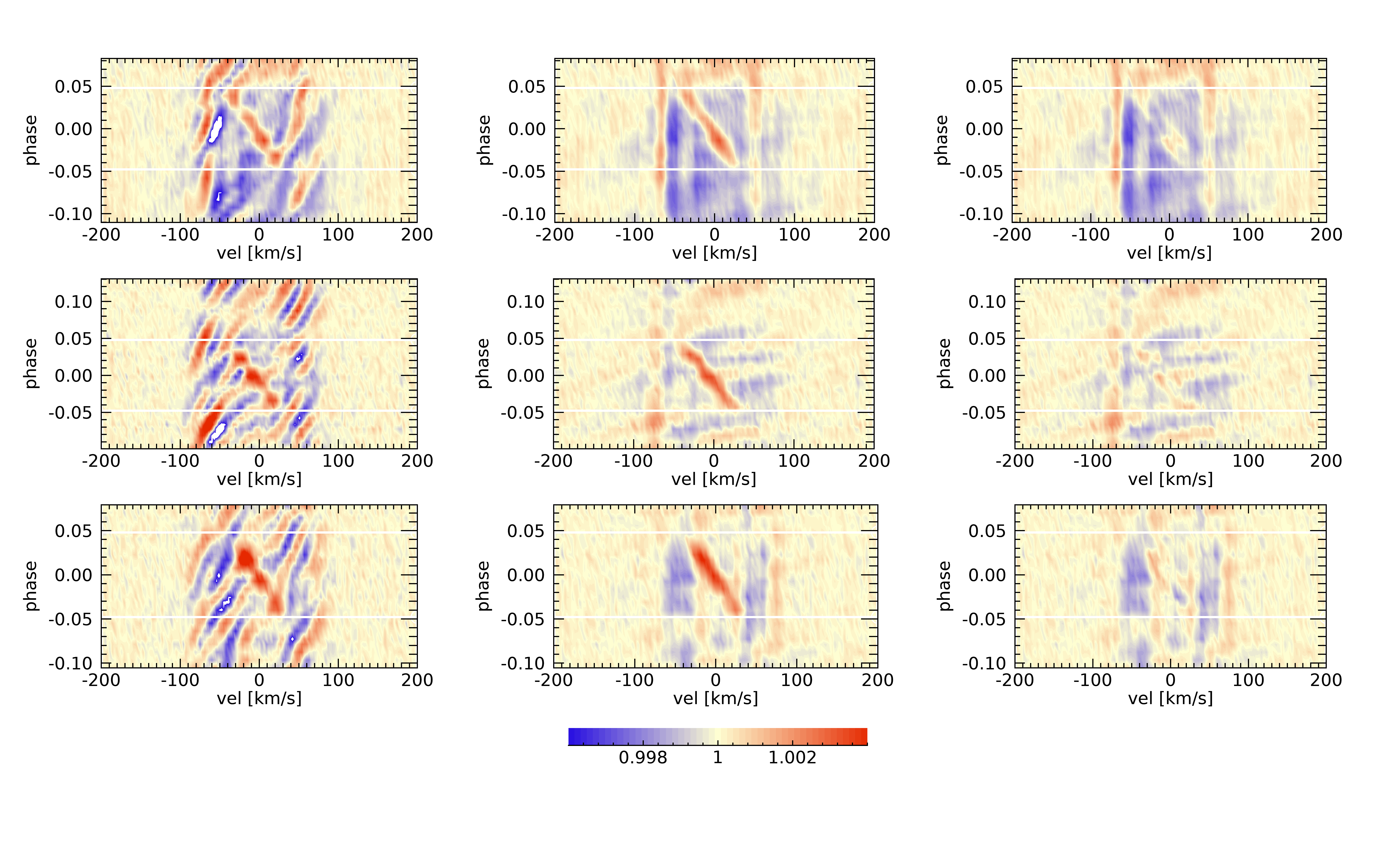}
\caption{Tomography of the pulsation filtering from the mean line profile residuals. From top to bottom:  transits 1, 3 and 4, respectively. {\it (Left panels)}: contour plot of the mean line profile residuals. {\it (Central panels)}: same as for left panels, but after filtering in the Fourier space. {\it (Right panels)}: same as for central panels, after subtracting the best fit Doppler shadow model. The horizontal white lines show the beginning and end of the transit. }
\label{fig:tomo_standard}
\end{figure*}

WASP-33 is a $\delta$-Sct A-type star. Since no default A-type mask is supported by the HARPS-N DRS pipeline \citep{2014SPIE.9147E..8CC}, we followed the approach presented in \citet{borsak9} and extracted the mean line profiles by means of the Least Square Deconvolution (LSD) software \citep{1997MNRAS.291..658D}. This software performs a Least-Squares Deconvolution of the normalised spectra with a theoretical line mask extracted from VALD \citep[Vienna Atomic Line Database, ][]{piskunov}. We used a stellar mask with T$_{\rm eff}$=7500 K, log$g$=4.0, and solar metallicity. 
We accurately re-normalised the spectra (order-by-order by using polynomials, see \citet{rainer2016} for a detailed description of the procedure) and converted them to the required format, working only on the wavelength regions $441.5-480.5$~nm, $491.5-528.5$~nm, $536.5-587.0$~nm, $605.0-626.5$~nm, and $633.5-645.0$~nm, i.e., cutting the blue orders where the signal-to-noise ratio (S/N) was very low due to the instrument efficiency, the Balmer lines, and the regions where most of the telluric lines are found. 
We then created mean line profile residuals by dividing all the mean line profiles by a master out-of-transit mean line profile for each transit observed.
As already evidenced by \citet{cameron} and \citet{Johnsonetal15}, the Doppler shadow of the planet is clearly visible as well as the stellar pulsations (Fig.~\ref{fig:tomo_standard}, left panel). 
We note that the pulsations are more evident on the edge of the lines with respect to the center, which is a hint of non-radial pulsations \citep[see also][]{cameron}.

For planets with projected spin-orbit angles close to 90 deg, nodal precession can be more easily detected during observations spanning several years \citep{Watanabeetal20}.
WASP-33b is one of the two reported cases where the exoplanet orbit has been found to show nodal precession \citep{Johnsonetal15,Watanabeetal20}, with the other case being Kepler-13Ab \citep{Szaboetal12,2018AJ....155...13H}.
Since our dataset extends the timespan of the reported orbital variations, we tried to see if this was visible also in our data.
We removed stellar pulsations independently for each single transit by using the Fourier transform filtering (Fig.~\ref{fig:tomo_standard}), following the method presented in \citet{Johnsonetal15}. This exploits the fact that pulsations (prograde) and Doppler shadow (retrograde) propagate in opposite directions: their frequency components thus tend to be separated in the two-dimensional Fourier transform of the line profile residuals time series \citep{Johnsonetal15}.
We then performed a fit to the Doppler shadow. The Doppler shadow model is taken from EXOFASTv2 \citep{2017ascl.soft10003E,2019arXiv190709480E}, passed through the Fourier filter and fitted to the data 
in a Bayesian framework by employing a differential evolution Markov chain Monte Carlo (DE-MCMC) technique \citep{TerBraak2006, Eastmanetal2013}, running five DE-MCMC chains of 100,000 steps and discarding the burn-in. 
We fixed \vsini, a/R$_{\rm s}$, T$_{0}$, period, R$_{\rm p}$/R$_{\rm s}$ to the values of Table~\ref{tab:parameters}. We left as free parameters the projected spin-orbit angle $\lambda$ and the inclination angle $i$, for which we set uninformative priors. 
The medians and the 15.86\% and 84.14\% quantiles of the posterior distributions were taken as the best values and $1\sigma$ uncertainties. 

\begin{figure}%[!ht]
\centering
\includegraphics[width=\linewidth]{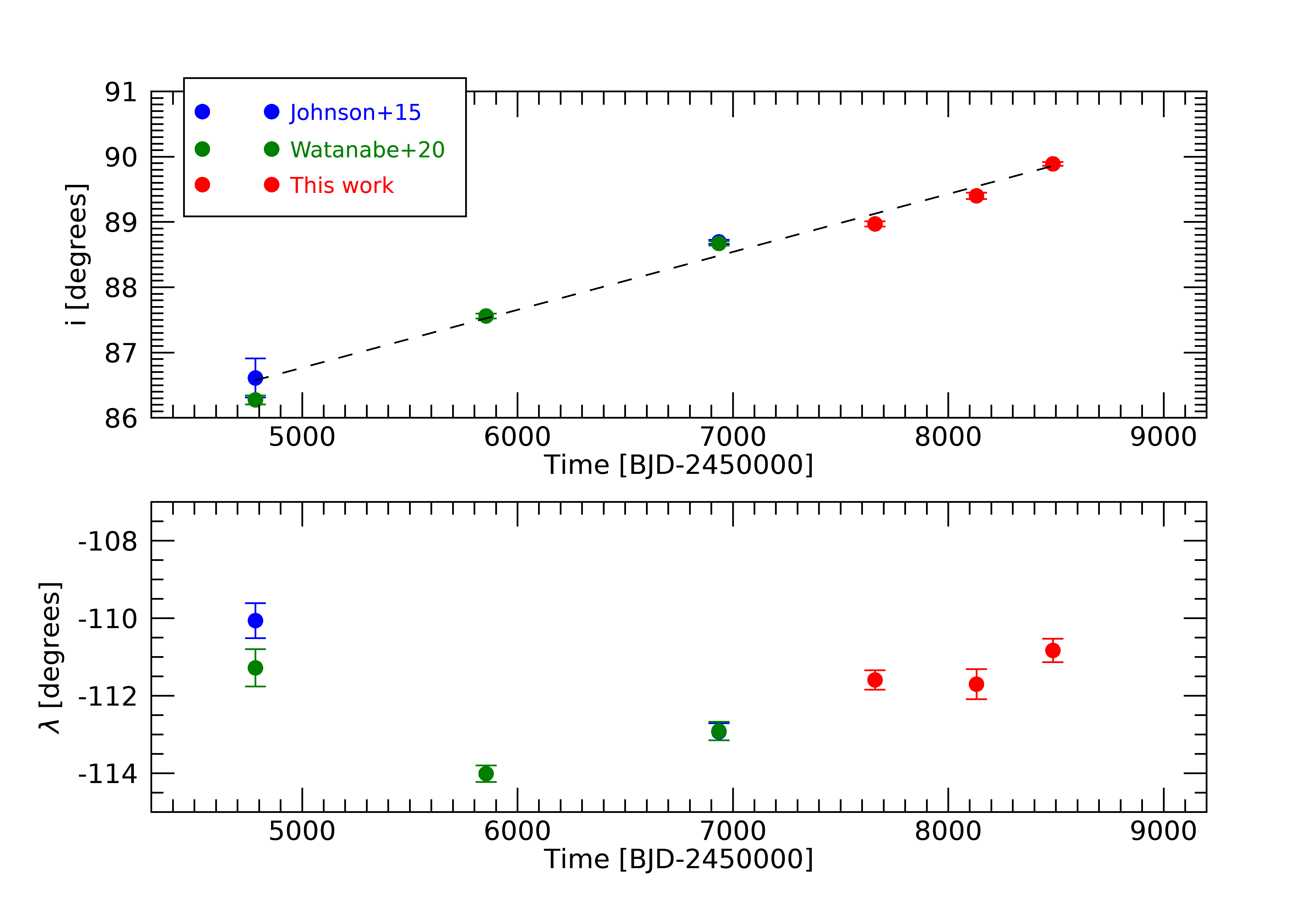}
\caption{Measurements of the inclination angle $i$ {\it (top panel)} and of the projected spin-orbit angle $\lambda$ {\it (bottom panel)} as a function of time. The dashed line shows the linear trend discussed in Sect. \ref{sec:mlp}.}
\label{fig:precession}
\end{figure}

Our results (Fig. \ref{fig:precession}, Table~\ref{tab:parameters}), together with the previous measurements of \citet{Watanabeetal20}, confirm the precession of the planetary orbit and that the obliquity projected on the plane of the sky $\lambda$ reached a minimum in 2011. This allows us to determine  the inclination of the stellar spin axis to the line of sight $i_{\rm s}$ and the obliquity $\epsilon$ at the epoch when $d\lambda/dt = 0$, thus allowing a more precise description of the precession of the system and the measure of the stellar gravitational quadrupole moment $J_{2}$. We describe the method applied to determine the system geometry as well as the rate of precession of the nodes of the orbital plane in Appendix~\ref{app_precession}. 

The rate of change of the inclination angle of the orbital plane $i$ can be regarded as constant over the time span of the observations ($\sim 10$ yr) because it is much shorter than the precession period ( $\sim 1100$~yr; see below). It is measured by a weighted linear best fit to the data in Fig.~2 (upper panel) giving $di/dt = 0.324 \pm 0.006$ deg/yr, while the epoch when $d\lambda/dt = 0$ is assumed to coincide with the transit observed on 19 October  2011, when $i=87.56^{\circ} \pm 0.037^{\circ}$ and $\lambda = -114.01^{\circ} \pm 0.22^{\circ}$ \citep{Watanabeetal20}. Given the uncertainty on such an epoch, we report the systematic deviations in the derived model parameters corresponding to a shift of  the epoch of $d\lambda/dt = 0$ by $\pm\, 500$ days since the assumed epoch, respectively (Table \ref{table_w33_angles}). To compute these deviations, we assume that $di/dt$ has the same value as above, while we linearly interpolate for the parameter $i$ and cubicly interpolate for the parameter $\lambda$ to find their values at the two epochs. We report for each parameter the standard deviation as a measure of the statistical error, while we add in brackets the systematic deviation when $d\lambda/dt=0$ at the later epoch and the deviation when $d\lambda/dt =0$ at the earlier epoch, respectively. Note that the two deviations have the same sign when the parameter reached an extremum in between the considered epoches.  

Following the method described in Appendix~\ref{app_precession}, we find that the stellar spin is perpendicular to the line of sight ($i_{\rm s} = 90.11^{\circ} \pm 0.12^{\circ} \; (-0.026^{\circ}; -0.015^{\circ})$), while the obliquity  $ \epsilon = 113.99^{\circ} \pm 0.22^{\circ} \; (-0.24^{\circ}; -0.67^{\circ})$ at the epoch of the transit on 19 October 2011. From the spectroscopic stellar $v\sin i_{\rm s}$, $i_{\rm s}$, and stellar radius, we estimate a rotation period of the star of $0.884 \pm 0.02$~days, not significantly affected by the systematic errors on $i_{\rm s}$ because it is close to $90^{\circ}$. The precession rate of the nodes of the orbital plane is found to be $-0.325 \pm 0.006 \; (6.7\times 10^{-5}; 1.1\times 10^{-4})$ deg/yr giving a precession period of $1108 \pm 19 \;(0.23; 0.38)$~yr, slightly longer than that determined by \citet{Johnsonetal15}. The time interval during which transits by WASP-33b are observable is thus of $\sim97$ yr centred around 2019, slightly longer than their interval of about $\sim88$ yr. 

Our inclination $i_{\rm s}$ of the stellar spin axis to the line of sight is different from the value found by \citet{2016MNRAS.455..207I} because his determination was based on the parameters derived only from the transits of 2008 and 2014 as reported by \citet{Johnsonetal15}. 
This led Iorio to consider a constant precession rate of the angle $I$ of his model, while our more extended dataset shows that $dI/dt$ is variable and became zero around 2011. In Appendix~\ref{appendix_iorio_model}, we account for the differences between his results and ours and show how the application of his model to our dataset reproduces our value of $i_{\rm s}$ and of the stellar quadrupole moment (see below).

Previous analyses of the precession of WASP-33b by \citet{Johnsonetal15} and \citet{Watanabeetal20} assumed that the stellar spin angular momentum is much larger than the orbital angular momentum because they adopted the gyration radius $\gamma$ typical of a Sun-like star following the exploratory calculations by \citet{Iorio11}. Here we determine more appropriate values of the gyration radius and apsidal motion constant $k_{2}$ of WASP-33 by making use of the tabulations of \citet{Claret19}. Considering the uncertainties in the stellar mass and effective temperature, we find that the Claret's model with  solar metallicity and  mass of $1.6$~M$_{\odot}$ is adequate to describe WASP-33. Taking into account the correction for its fast rotation, we find $\log k_{2} = -2.55 \pm 0.014$ and $\gamma = 0.1884 \pm 0.0019$,  where the uncertainties take into account only the uncertainty in the stellar effective temperature. 

With the above values of $k_{2}$, $\gamma$, and the adopted stellar and planetary parameters, we predict a value of $di/dt = 0.315 \pm 0.026 \; (0.0023; 0.0068)$ deg/yr and a precession rate of the nodes of $-0.316 \pm 0.026 \; (0.0024; 0.0069)$ deg/yr using  the formulae given in Appendix~\ref{app_prec_formula}. The coincidence of these numerical values comes from the geometry of the system as explained there. These precession rates differ by less than a half standard deviation from the observed ones, supporting the correctness of the adopted stellar parameters. With our value of $\gamma$ and the adopted system parameters, we find that the ratio of the stellar spin to the orbital angular momentum is $2.75 \pm 0.33$, therefore, we cannot neglect the precession of the stellar spin. It occurs with the same period as the precession of the node of the orbital plane and makes the inclination $i_{\rm s}$ of the stellar spin axis to the line of sight vary between $67.5^{\circ}$ and $110.8^{\circ}$ along a complete precession cycle. Note that an inclination of the stellar spin greater than $90^{\circ}$ implies that the South pole of the star is in view, so the apparent stellar rotation is clockwise, contrary to the usual anticlockwise rotation assumed when $ i_{\rm s} < 90^{\circ}$ and the North pole is in view. 

The fast stellar rotation produces a deformation in the stellar mass distribution and hence a distortion in the gravitational field.
The stellar gravitational quadrupole moment $J_{2}$ can be deduced from the observed nodal precession rate of the orbital plane according to eq.~(3) of \citet{Johnsonetal15} and is $(6.73 \pm 0.22 \; [0.062; 0.18]) \times 10^{-5}$. This value compares well with the theoretically expected value of $(6.53 \pm 0.41\; [0.011; 0.031]) \times 10^{-5}$ \citep[e.g.,][and Appendix~\ref{app_quad_mom}]{RagozzineWolf09}, thus confirming the goodness of the adopted values of $k_{2}$ and $\gamma$. It also supports the hypothesis that WASP-33b is the only close massive planet in the system because another similar body would contribute to the orbital precession rate, if its orbit were not coplanar with that of WASP-33b. The fast rotation of WASP-33 makes its quadrupole moment $\sim 375$ times larger than the solar value \citep[$J_{2\odot} \sim 1.8 \times 10^{-7}$;][]{Rozelotetal09}, the effect of the centrifugal force being only modestly compensated by the stronger density concentration in an A-type star.

\begin{table}
\begin{center}
\caption{Stellar and orbital parameters adopted in this work.}
\label{tab:parameters}
\footnotesize
\begin{tabular}{ccc}
 \hline\hline
 \noalign{\smallskip}
 Parameter & Value & Reference\\
 \noalign{\smallskip}
 \hline
\noalign{\smallskip}
\multicolumn{3}{c}{WASP-33}\\
\noalign{\smallskip}
\hline
\noalign{\smallskip}
\multicolumn{3}{c}{\it Stellar parameters}\\
\noalign{\smallskip}
 $T_{\rm eff}$ [K] &   7430 $\pm$ 100 & Yan et al. 2019 \\
log g &  4.3 $\pm$ 0.2 & Yan et al. 2019\\
 $[Fe/H]$ & -0.1 $\pm$ 0.2 & Yan et al. 2019\\
 $R_{\rm s}$ [R$_{\odot}$]& 1.509 $\pm$ 0.025 &  Yan et al. 2019\\
 $M_{\rm s}$ [M$_{\odot}$] & 1.561 $\pm$ 0.06 &Yan et al. 2019  \\
\vsini \ [\kms] &  86.4 $\pm$ 0.5 & This work\\
$P_{\rm rot}$ [days] & 0.884 $\pm$ 0.02 &This work \\
\noalign{\smallskip}
\multicolumn{3}{c}{\it Orbital parameters}\\
\noalign{\smallskip}
Period [days]&   $1.219870897$ & Yan et al. 2019 \\
$T_0$ [BJD-2450000] &   $4163.22449$ & Yan et al. 2019 \\  
$R_{\rm p}$/$R_{\rm s}$ & $0.11177$   & Yan et al. 2019 \\
$a/R_{\rm s}$ &  $3.69\pm0.05$  & Yan et al. 2019 \\
$e$ &   $0.0$ & assumed\\
$i$ [degrees]&   $88.97\pm0.04$ & This work, transit 1\\
$i$ [degrees]&   $89.40\pm0.05$ & This work, transit 3\\
$i$ [degrees]&   $89.89\pm0.03$ & This work, transit 4\\
$\lambda$ [degrees]&   $-111.59\pm0.25$ & This work, transit 1\\
$\lambda$ [degrees]&   $-111.70\pm0.39$ & This work, transit 3\\
$\lambda$ [degrees]&   $-110.83\pm0.30$ & This work, transit 4\\
V$_{\rm sys}$ [\kms] & $-2.76\pm0.04$ & This work\\
K$_{\rm p}$ [\kms] & $231\pm3$ & Yan et al. 2019 \\
M$_{\rm p}$ [M$_{\rm Jup}$] & $2.16\pm0.20$ & Yan et al. 2019 \\
T$_{\rm eq}$ [K] &2710 $\pm$ 50 & Yan et al. 2019 \\
\noalign{\smallskip}
 \hline
\end{tabular}
\end{center}
\end{table}

%__________________________________________________________________

\section{Radial velocities\label{sec:rv}}

\begin{figure*}%[!ht]
\centering
\includegraphics[width=\linewidth]{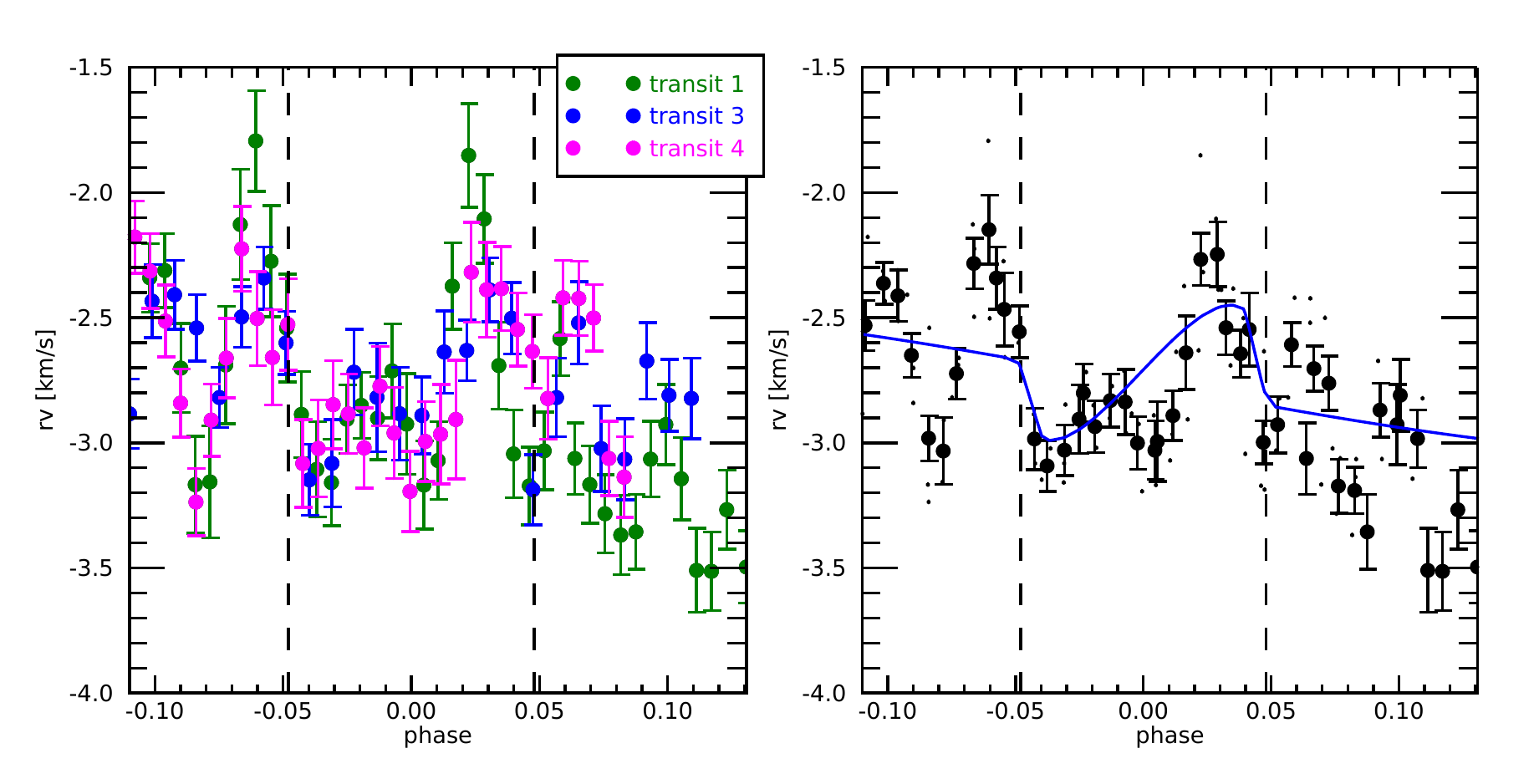}
\caption{{\it (Left panel)} Phase folded RVs of the three analysed HARPS-N transits of WASP-33b. Different colours refer to different transits. {\it (Right panel)} RVs of the three transits averaged in bins of 0.005 in phase (filled circles), together with single RVs (small dots). The blue line represents the theoretical RV solution calculated with the parameters of Table \ref{tab:parameters} (the values of $\lambda$ and $i$  are obtained from the average of the three transits).}
\label{fig:rml_doppio}
\end{figure*}

Since we could not exploit the Cross-Correlation Functions (CCFs) from the DRS, RVs were extracted following the same method presented in \citet{borsak9}.
Instead of using a Gaussian fit, we preferred to model the LSD lines with a rotational profile \citep{gray}. 
We measured a \vsini \ =$86.4\pm0.5$ \kms, taken as the average of the out-of-transit measurements and their standard deviation, which is well in agreement with previous values \citep[\vsini \ =$86.63$ \kms,][]{Johnsonetal15}.
The extracted RVs are listed in Table \ref{tab:app_rv}. 
We subtracted from each transit observation the difference from the mean of the in-transit RVs, to avoid any possible offset caused by instabilities, long-term pulsations and trends in the orbital solution \citep[e.g.,][]{borsak9}, and then averaged the RVs of the four transits in bins of 0.005 in phase.
The RV time series of each single transit analysed and their average are shown in Fig.~\ref{fig:rml_doppio}.

Although stellar pulsations dominate also the RV curve, it is evident a qualitative agreement with the theoretical RM curve calculated using the parameters presented in Table~\ref{tab:parameters} \citep[by using the formalism of][]{ohta}.
This is a further confirmation that for fast rotators the method of fitting the mean line profiles with a rotational profile instead of a Gaussian function brings optimal results, as it was previously found for other targets \citep[e.g.,][]{2018arXiv180904897A,2018AJ....155..100J,borsak9,rainerk20}. 
We note however that for this case the Doppler tomography method remains the preferred one to determine the projected spin-orbit inclination angle.

%\begin{table}
%\begin{center}
%\caption{HARPS-N RV observations of WASP-33. This table is available in its entirety online at the CDS.}
%\label{tab:rv}
%\footnotesize
%\begin{tabular}{ccc}
% \hline\hline
% \noalign{\smallskip}
%Time [BJD-2450000] & RV [\kms] & RV error [\kms]\\
% \noalign{\smallskip}
% \hline
%\noalign{\smallskip}
%7660.46966   &   -2.210  &   0.137 \\
%7660.47683  & -2.181 & 0.148 \\
%7660.48443 & -2.571 & 0.178 \\
% 7660.49131 & -3.037 & 0.194\\
% ... & ...  & ...\\
%\noalign{\smallskip}
% \hline
%\end{tabular}
%\end{center}
%\end{table}

%______________________________________________________________
\section{Balmer lines absorption in the transmission spectrum\label{sec:Halpha_main}}

We studied the atmosphere of WASP-33b by using the technique of transmission spectroscopy, focusing on H$\alpha$ and H$\beta$ absorption and taking into account the possible contamination given by stellar effects.

\subsection{Transmission spectrum extraction\label{sec:trans_spectrum}}

We performed transmission spectroscopy following the method of \citet{wyttenbach}, applying the following steps independently to each transit.
First we shifted the spectra to the stellar radial velocity rest frame using a Keplerian model of the system.
Then we normalized each spectrum, by dividing for the average flux within defined wavelength ranges where telluric and stellar lines are not present.

Using the out-of-transit spectra only, we built a telluric reference spectrum $T(\lambda)$, by means of  
a linear correlation between the logarithm of the normalized flux and the airmass \citep[][]{2008A&A...487..357S,2010A&A...523A..57V,2013A&A...557A..56A}, 
and then rescaled all the spectra as if they had been observed at the airmass corresponding to that at the center of the transit. 
 We then created a master stellar spectrum $S_{\rm master}$ by averaging all the out-of-transit spectra (excluding the phase range of the immediate pre-transit, see Sect. \ref{sec:pretransit}), and divided all the single spectra for this $S_{\rm master}$ creating the residual spectra $S_{\rm res}$.
Every $S_{\rm res}$ was then shifted considering the theoretical planetary RV and the systemic velocity of the system, i.e., we placed the spectra in the planetary reference frame. Here we expect to detect the exoplanetary atmospheric signal centered at the laboratory wavelengths. All the full-in-transit $S_{\rm res}$ were then averaged to create the transmission spectrum.
At this stage, the transmission spectrum of the planet still includes spurious stellar contaminations. 

\subsection{Removal of CLV + RM\label{sec:clv}}
The star over which the planet transits is not a simple homogeneous disk, but rotates and has a surface brightness which changes as a function of the distance from center.
Effects such as center-to-limb variations (CLV) and RM have been proven to significantly modify the shape of line profiles, possibly causing false atmospheric detections \citep[e.g.,][]{yan,bez,2020arXiv200210595C}. 
We thus took these effects into account by creating a model following the methodology described in \citet{yan}. 
The star is modeled as a disk divided in sections of 0.01 \rs. For each point, we calculate the $\mu$ value (where $\mu=\cos{\theta}$, with $\theta$ the angle between the normal to the stellar surface and the line of sight) and the projected rotational velocity (by assuming rigid body rotation and rescaling the \vsini \ value of Table \ref{tab:parameters}).
A spectrum is then assigned to each point of the grid, by quadratically interpolating on $\mu$ and Doppler-shifting for the stellar rotation the model spectra created using the tool Spectroscopy Made Easy \citep[SME,][]{2017A&A...597A..16P}, with the line list from the VALD database \citep{2015PhyS...90e4005R} and the MARCS \citep{2008A&A...486..951G} stellar atmospheric models (assuming local thermodynamic equilibrium approximation). The model spectra are created with null rotational velocity for 21 different $\mu$ values, and adapted to the resolving power of the instrument.
Then, using the orbital information from Table \ref{tab:parameters} we simulate the transit of the planet, calculating the stellar spectrum for different orbital phases as the average spectrum of the non-occulted modeled sections. As the last step, we divide each spectrum for a master stellar spectrum calculated out-of-transit, obtaining the information of the impact of RM+CLV effects at each in-transit orbital phase.
We then move everything in the planetary restframe and calculate the simulated CLV+RM effects on the transmission spectrum (Fig. \ref{fig:corrections}). 
We note that we do not take into account the possible deformation of the line profiles given by gravity darkening, which are however not expected to be significant \citep[see, e.g.,][]{Kovacsetal13}.

\subsection{Removal of pulsations\label{sec:puls}}

\begin{figure}%[!ht]
\centering
\includegraphics[width=\linewidth]{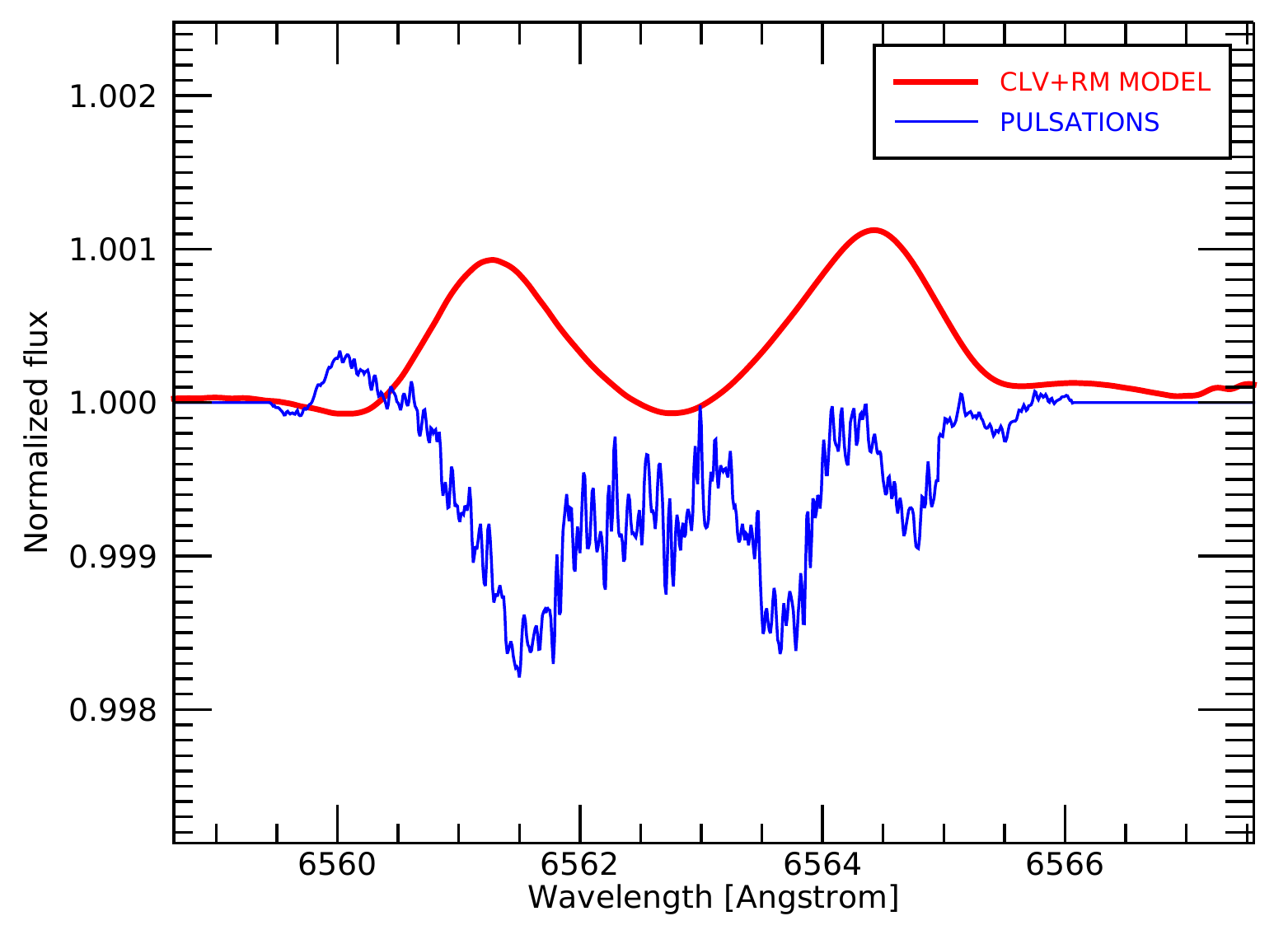}
\caption{The corrections for CLV+RM and for stellar pulsations applied to the transmission spectrum in the H$\alpha$ zone.} 
\label{fig:corrections}
\end{figure}

\begin{figure*}%[!ht]
\centering
\includegraphics[width=\linewidth]{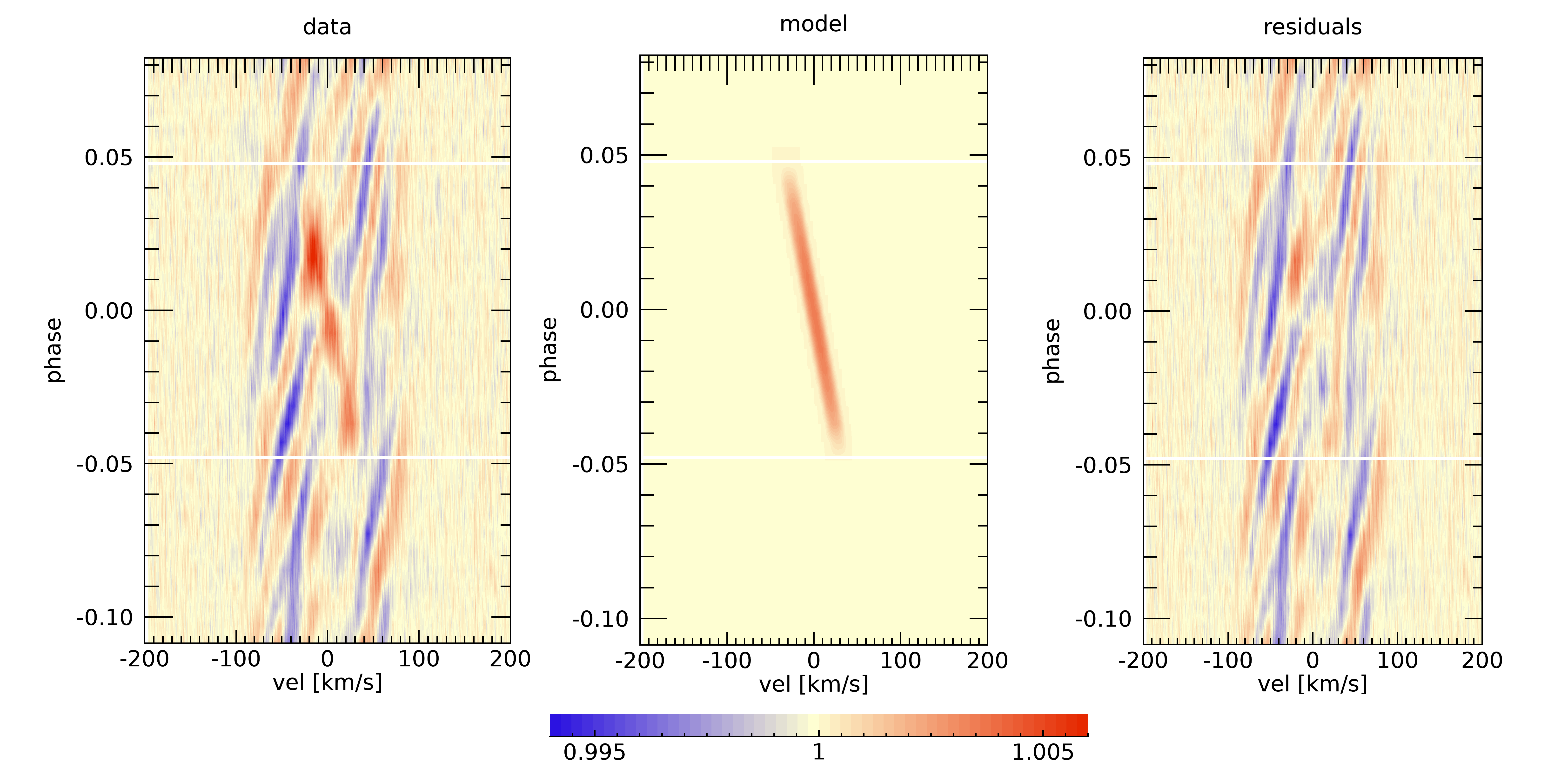}
\caption{Example of the removal of the Doppler shadow from the mean line profile residuals for transit 4. {\it (Left panel)} Original mean line profile residuals. {\it (Central panel)} Model of the Doppler shadow. {\it (Right panel)} Mean line profile residuals after the removal of the Doppler shadow. The horizontal white lines show the beginning and end of the transit.}
\label{fig:triplo_tomo}
\end{figure*}

\begin{figure}%[!ht]
\centering
\includegraphics[width=\linewidth]{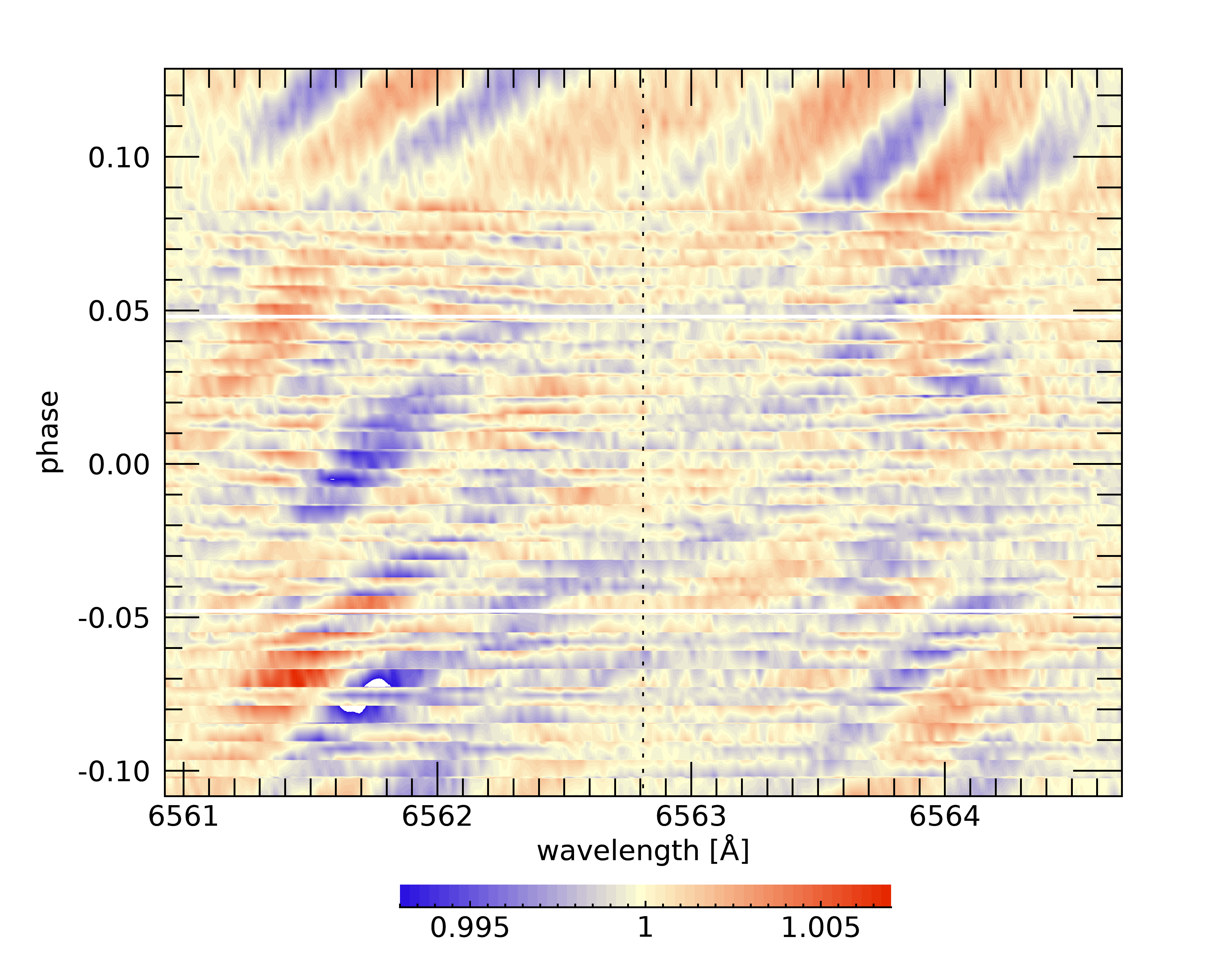}
\caption{The effect of the pulsations contamination on the three analysed transits as it affects the H$\alpha$ line, in the stellar restframe. The vertical dotted line marks the position of the H$\alpha$ line. The horizontal white lines show the beginning and end of the transit.}
\label{fig:pulsations_tomo}
\end{figure}

Another contamination one has to deal with in this particular case of WASP-33b comes from the stellar pulsations. 
We already showed in Sect. \ref{sec:mlp} how strongly they impact the mean line profiles. In the same way, they will have an impact in the extracted planetary transmission spectrum.
To deal with this, we used an approach similar to that presented in \citet{bez} to remove the RM from the transmission spectrum by using real observing data. This was done by exploiting the S/N of the mean line profiles.
We note that pulsations and RM cannot be removed together in this way. This is because the amplitude of the RM effect is dependent on the planetary  effective radius in the narrow bandpasses where the planetary lines are present. On the contrary, the amplitude of pulsations vary on a much larger wavelength range and can be assumed constant in each of our narrow bandpasses of interest.
In the following, we make the assumption that the pulsation pattern behaves in the same way for all the absorption lines. We use the pulsation pattern extracted from the mean line profile (which is basically based on metal lines) and assume that it is valid also for the Balmer lines, except for a scaling factor. 
To justify this assumption we note that the short periods of $\delta$ Sct are not able to produce significant phase shifts between the photospheric stellar lines of different species and excitation potentials. Moreover, the broadening of the Balmer lines largely affects the wings, but not so much the center, where we search for the in-transit planetary RVs: Balmer and metallic line-profiles are very similar
there. We also note that we cannot analyse the pulsation pattern directly on the Balmer lines, due to their low S/N. As a final test we verified that the applied correction does not impact the depth of the retrieved planetary line-profile (Sect. \ref{sec:halpha}).

We started from the mean line profiles residuals, after the division for an average out-of-transit mean line profile. We then subtracted for each transit the Doppler shadow models calculated in Sect.~\ref{sec:mlp} (e.g., Fig.~\ref{fig:triplo_tomo}). Then we created a transmission spectrum of the pulsations (TS$_{puls}$) in the wavelength region of interest. First we put together the three transits, then we move from the velocity to the wavelength space, taking as zero reference the laboratory wavelength of the H$\alpha$ (Fig. \ref{fig:pulsations_tomo}) and H$\beta$ lines, since these are the lines we aim at correcting.
Then we moved each mean line profile residual to the planetary restframe, and averaged the full-in-transit residuals to calculate the TS$_{puls}$.
At this point we still have to correct for the magnitude of the effect: the calculated TS$_{puls}$ is in fact referred to a line whose depth is the one of the mean line profile, and has to be rescaled to the depth of the line that we want to correct \citep{bez}.
We thus multiplied the amplitude of the correction for a factor which is the ratio between the depth of the mean line profile and that of the H$\alpha$ (or H$\beta$) line, and obtained the final spectrum used to correct for pulsations (Fig. \ref{fig:corrections}).

\subsection{Balmer lines absorption\label{sec:halpha}}

\begin{figure}%[!ht]
\centering
\includegraphics[width=\linewidth]{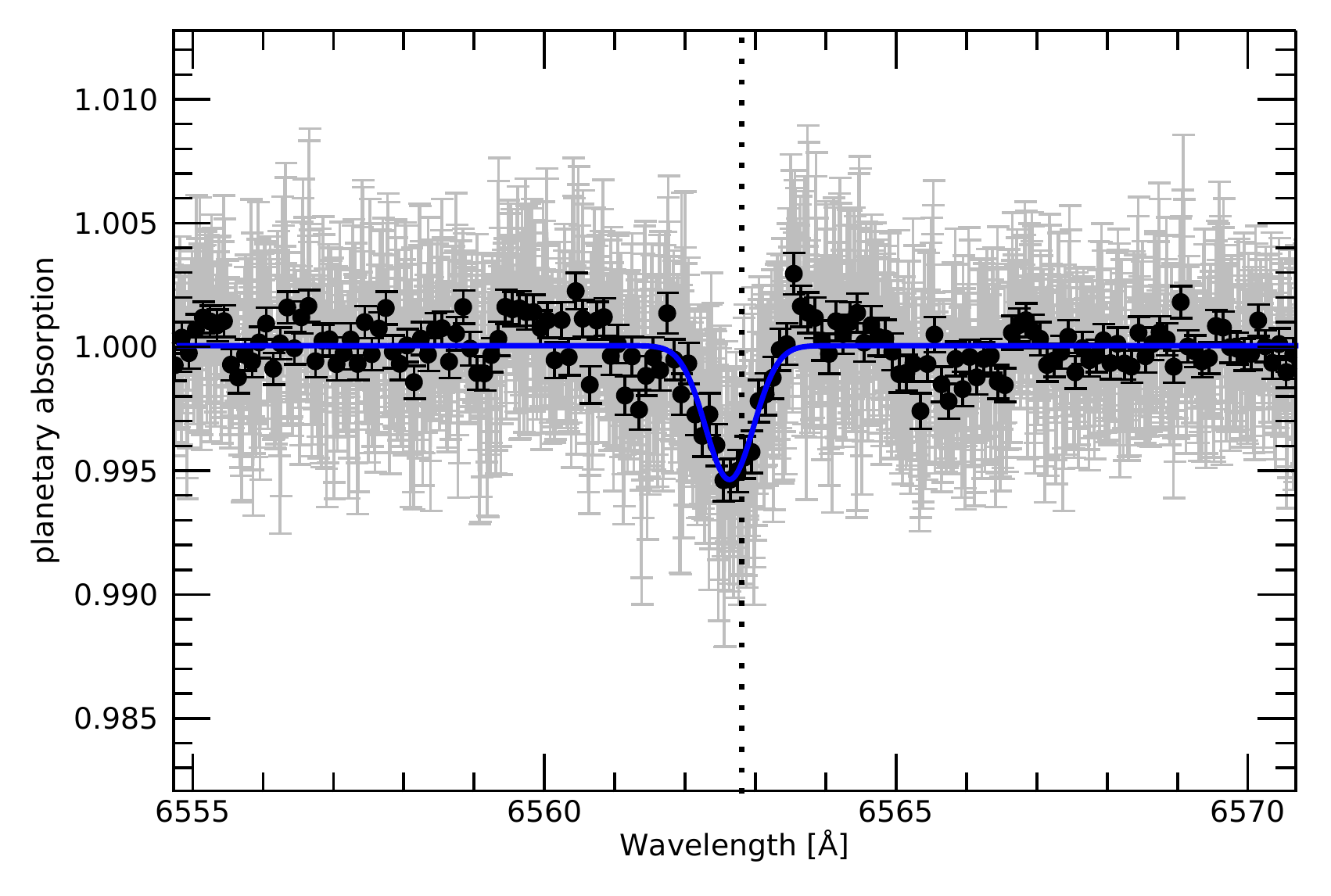}
\caption{H$\alpha$ absorption after the correction for stellar effects (RM+CLV and pulsations). Black circles represent 0.1 \AA\ binning. The blue line is the Gaussian best fit. The vertical dotted line shows the planetary H$\alpha$ restframe.}
\label{fig:Halpha}
\end{figure}

We corrected the transmission spectrum extracted in Sect. \ref{sec:trans_spectrum} from stellar effects by dividing it for the corrective spectra calculated in Sect. \ref{sec:clv} and \ref{sec:puls}. We note that the effect of pulsations overcomes that of CLV+RM by a factor $\sim$1.5 in magnitude (Fig. \ref{fig:corrections}).
We find an excess absorption in the H$\alpha$ region with a contrast of 0.54 $\pm$ 0.04\% and a FWHM of 36.7 $\pm$ 3.3 \kms\ (Fig.~\ref{fig:Halpha}). We can translate the contrast into an effective planetary radius R$_{\rm eff}$=1.18 $\pm$ 0.02 R$_{\rm p}$, calculated assuming $R^2_{\rm eff}/ R^2_{\rm p}=(\delta+h)/\delta$, with $\delta$ the transit depth (from Table \ref{tab:parameters}) and $h$ the line contrast \citep[e.g.,][]{2020A&A...635A.171C}.
As for KELT-9b, the H$\alpha$ profile of WASP-33b forms in the atmospheric region below the Roche lobe, which lies at about 1.6 planetary radii. Therefore this detection does not directly probe atmospheric escape, but it would enable one to set constraints on the planetary temperature-pressure structure and thus on the energetics possibly driving escape.
The line has a significant blueshift of -8.2 $\pm$ 1.4 \kms, indicative of winds in the planetary atmosphere. 
We note that without the correction for the stellar pulsations we find a comparable depth (0.53\%) but a larger FWHM (55 \kms). Being non-radial pulsations, their averaged effect on the three transits tends to be smaller in the center of the stellar line profile, and this is reflected also in the transmission spectrum. This is also noticeable in Fig. \ref{fig:corrections} and Fig. \ref{fig:pulsations_tomo}.

\begin{figure}%[!ht]
\centering
\includegraphics[width=\linewidth]{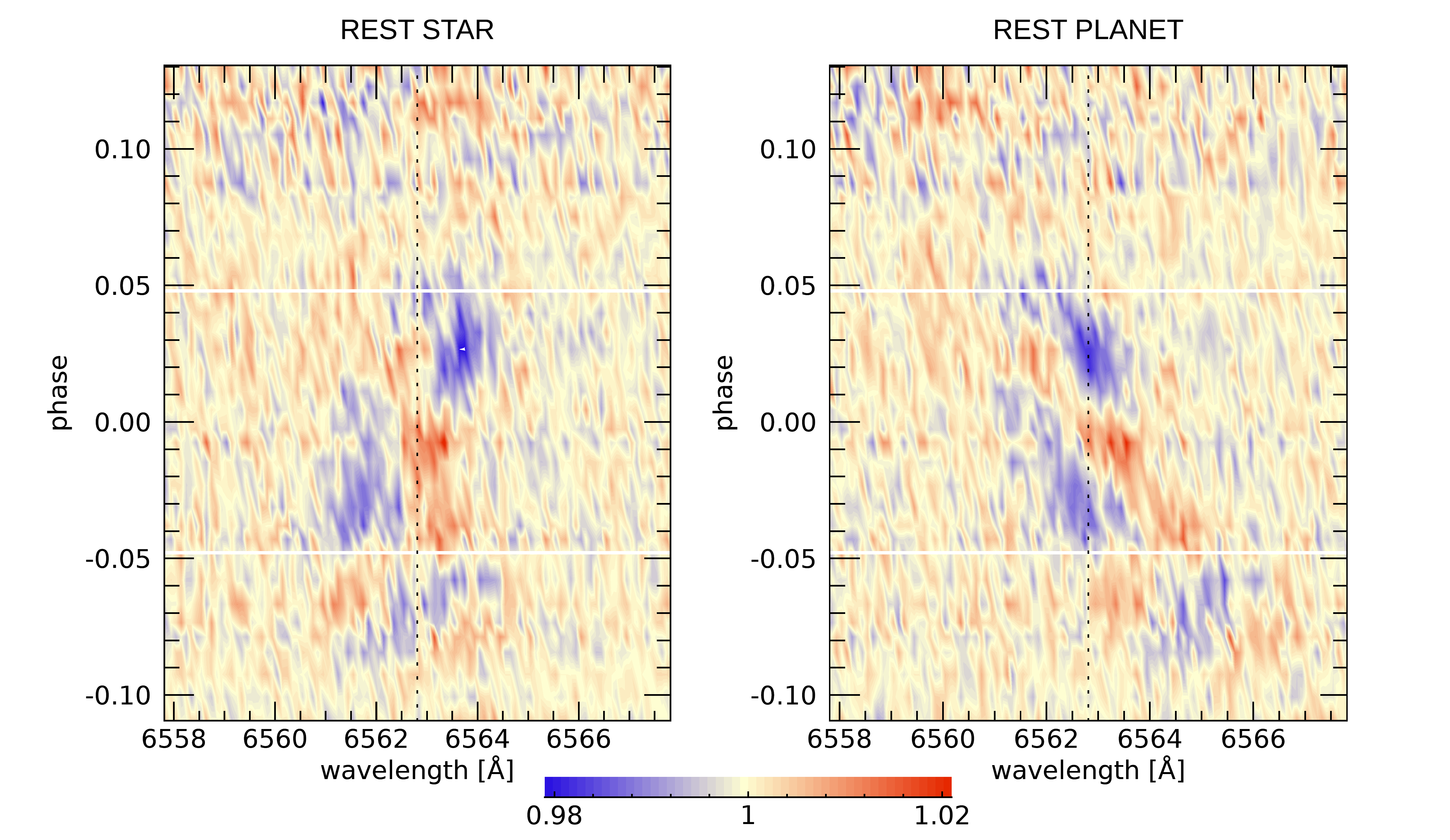}
\caption{Contour 2D tomography map of H$\alpha$ absorption in the stellar (left panel) and planetary (right panel) restframes, before applying the stellar contamination correction (this evidences also the  Doppler shadow, i.e., the red track). The white horizontal lines represent the beginning and end of the transit.
The vertical dotted black line shows the H$\alpha$ planetary restframe in the right panel and the stellar restframe in the left panel. A pre-transit signal, not centered in the planetary restframe, is evident (see discussion in Sect. \ref{sec:pretransit}).}
\label{fig:tomo_Halpha}
\end{figure}

Checking the reference frame of the detections is fundamental when we want to be sure they are caused by the planetary atmosphere and not by spurious stellar effects \citep[e.g.,][]{2016ApJ...817..106B,bez}, in particular for this case where the strong stellar pulsations could be not perfectly corrected for and mimic atmospheric features. We thus verified that our H$\alpha$ absorption detection is in the planetary restframe by resolving it in the 2D tomographic map (Fig.~\ref{fig:tomo_Halpha}), which shows the position of the absorption signal as a function of the orbital phase.

\begin{figure}%[!ht]
\centering
\includegraphics[width=\linewidth]{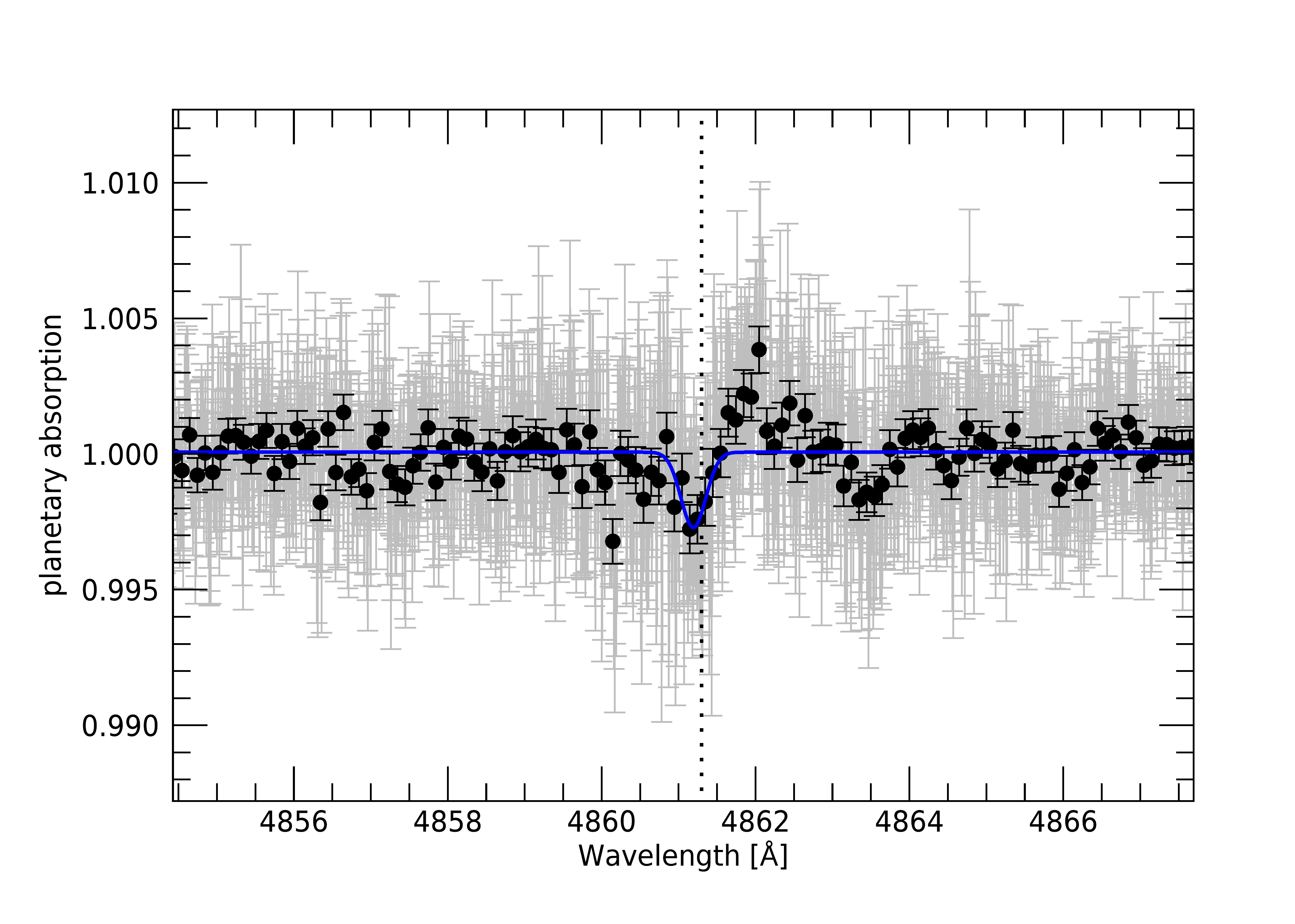}
\caption{Same of Fig. \ref{fig:Halpha} but for H$\beta$.}
\label{fig:Hbeta}
\end{figure}

\begin{figure}%[!ht]
\centering
\includegraphics[width=\linewidth]{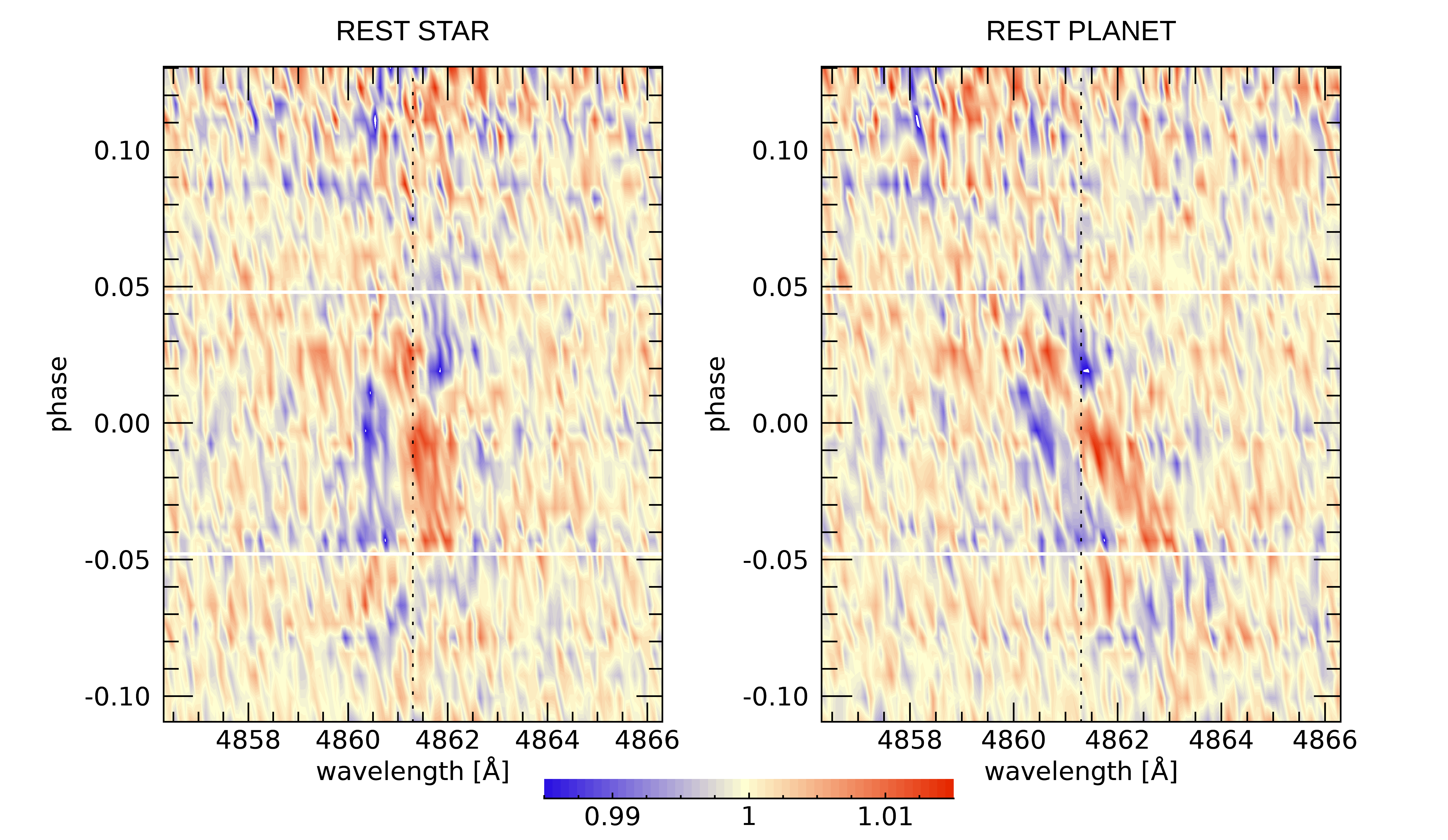}
\caption{Same of Fig. \ref{fig:tomo_Halpha} but for H$\beta$.}
\label{fig:tomo_hbeta}
\end{figure}

We also find an excess absorption in the H$\beta$ line region with a contrast of 0.28$\pm$0.06\%, FWHM of 23.5$\pm$6.3 \kms\ and a blueshift of -6.6$\pm$2.6 \kms\ (Fig. \ref{fig:Hbeta} and Fig. \ref{fig:tomo_hbeta}). We can translate the contrast into an effective planetary radius R$_{\rm eff}$=1.09$\pm$0.03 R$_{\rm p}$.

We note that the values of absorption depth we obtain are almost half of those obtained for the same planet by \citet{2020arXiv201107888Y} (0.99$\pm$0.05\% for H$\alpha$ and 0.54$\pm$0.07\% for H$\beta$, respectively). However, the H$\alpha$/H$\beta$ line depth ratio is the same (1.93$\pm$0.44 for us versus 1.83$\pm$0.26 for them). 
Given the fact that also another analysis of these lines shows different line profile depths \citep[1.68$\pm$0.02\% for H$\alpha$ and 1.02$\pm$0.05\% for H$\beta$,][]{cauley2020}, we highlight the possibility that the amplitude of the lines could be variable with time and with the level of the stellar activity \citep[e.g.,][]{borsaespresso} and/or pulsations. 
We checked within the dataset we analysed, without finding evidence of significant variability with all the three transits having the same H$\alpha$ contrast.

A future common analysis of all available datasets will help shedding light on the possibility of the observed transit depth variations being related to stellar activity and/or pulsation, rather than caused by systematics between different analyses. Furthermore, such an analysis will enable one to obtain a very high-quality average line profile for each detected Balmer line. These profiles can then be used to constrain the temperature structure of the planetary atmosphere in the 10$^{-3}$--10$^{-9}$\,bar range \citep[e.g.,][]{2020A&A...643A.131F}, going beyond the analysis of \citet{2020arXiv201107888Y}, who employed an isothermal profile and assumed local thermodynamical equilibrium. Indeed, it has been shown that, for modelling the Balmer lines detected in the transmission spectrum of the UHJ KELT-9b, these assumptions are invalid \citep{2020A&A...643A.131F}. Because of similarities in the temperature profile of the upper atmosphere of KELT-9b and WASP-33b \citep[e.g., both present a thermal inversion in the upper atmosphere and reach upper atmospheric temperatures of the order of 10000\,K;][]{2018ApJ...868L..30F,2020A&A...643A.131F}, it is likely that the assumptions employed by \citet{2020arXiv201107888Y} for modelling the Balmer lines led them to obtain unreliable results.

%______________________________________________________________
\section{Pre-transit signal\label{sec:pretransit}}

\begin{figure*}%[!ht]
\centering
\includegraphics[width=\linewidth]{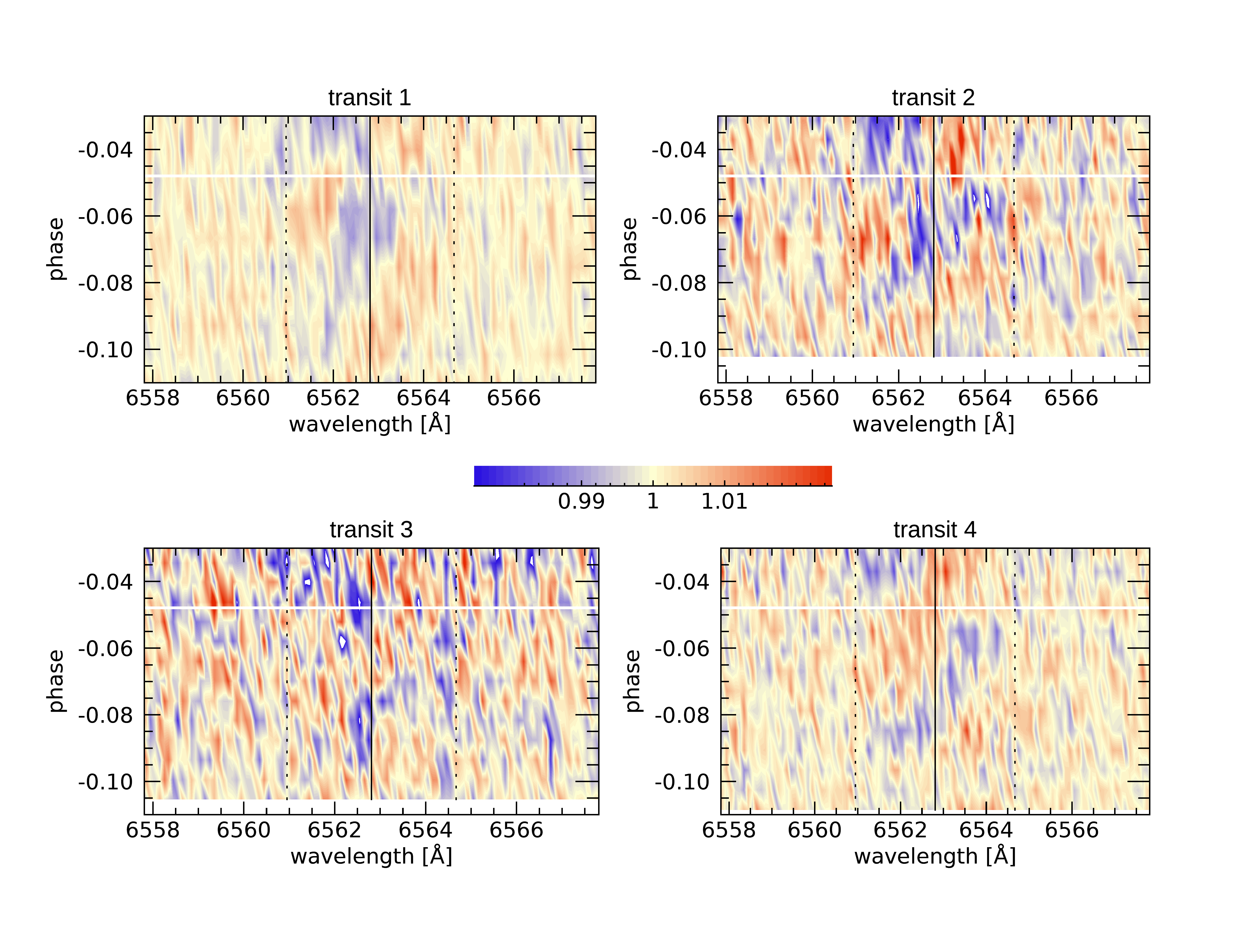}
\caption{Zoom on the pre-transit signal on the 4 transits of our dataset, in the stellar restframe. Vertical line shows the H$\alpha$ line center, while vertical dotted lines mark the $\pm$\vsini\ limits. Horizontal white line shows the beginning of the transit. The pre-transit signal is the diagonal blue track before the beginning of the transit.}
\label{fig:4transits}
\end{figure*}

\begin{figure}%[!ht]
\centering
\includegraphics[width=\linewidth]{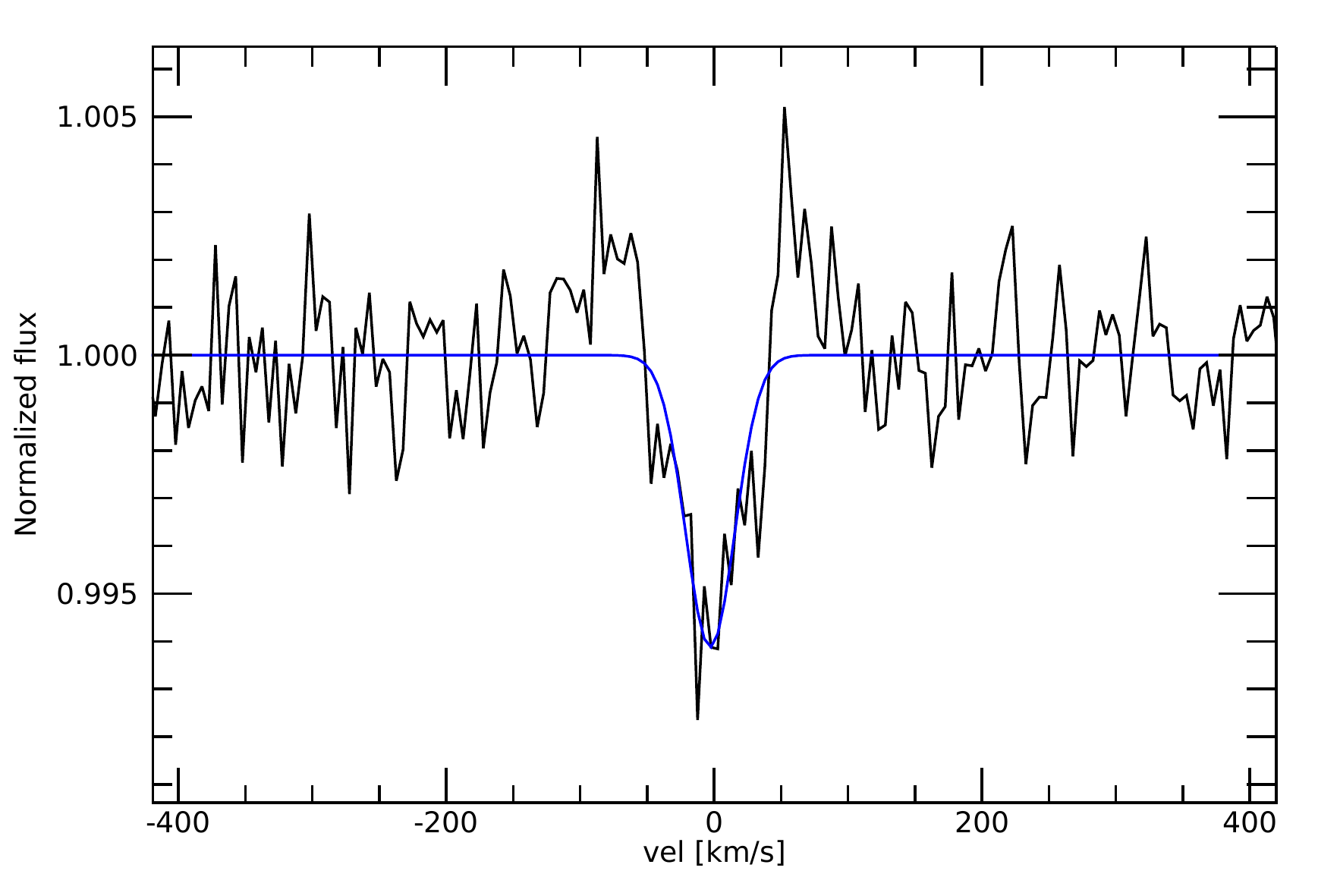}
\caption{Pre-transit signal profile in its restframe with K$_{\rm PTS}$=460 \kms. The blue line shows the best-fit Gaussian profile.}
\label{fig:pts_profile}
\end{figure}

Looking at the tomographic map of the H$\alpha$ absorption (Fig.~\ref{fig:tomo_Halpha}), we noted a pre-transit feature.
This feature has a different slope and duration in the 2D tomographic map with respect to the dominant pulsation pattern in the mean line profiles (see Fig.~\ref{fig:tomo_standard} and Fig.~\ref{fig:4transits}). 
We checked one by one the single transits, and found that this feature looks quite similar in all four available transits (Fig.~\ref{fig:4transits}, see the blue diagonal track before the beginning of the transit). It is thus something which happens in an almost coherent timing with the orbital period of the planet.

It appears to be a pseudo-absorption pre-transit signal (PTS hereafter) happening on the stellar surface, because it is moving from -\vsini\ to +\vsini\ (i.e., across the stellar CCF) with time.
We thus investigated this more in detail, performing the normalization on an extended out-of-transit baseline to avoid including the PTS in it.
By visual inspection, we noted that a similar behaviour is observed in many single lines, not only on H$\alpha$. On the mean line profiles it is hidden by the other stellar pulsations, but is still noticeable after removing them (Fig. \ref{fig:tomo_standard}, right panel). Curiously, this feature is also reflected in the radial velocities (see Fig. \ref{fig:rml_doppio}).

We found that the restframe of the PTS is not compatible with the one of the planet, by studying it on the H$\alpha$ line and assuming it is moving with a Keplerian motion around the star with the same orbital period of the planet. Thus we define K$_{\rm PTS}$ as the Keplerian semi-amplitude of this pseudo-orbital motion.
By fitting Gaussian functions and maximizing the contrast of their average, we could estimate that the PTS is moving at a K$_{\rm PTS}\sim$ 460 \kms\ and is centered at phase $\sim$-0.07. As averaged on these values, the PTS has a contrast of 0.61$\pm$0.06 \% (Fig. \ref{fig:pts_profile}). 
Curiously, the value of the K$_{\rm PTS}$ is almost double as the one of K$_{\rm p}$ (231\kms, Table \ref{tab:parameters}).
The two shoulders around the profile (Fig. \ref{fig:pts_profile}) and the red tracks around the signal (Fig. \ref{fig:4transits}) mimic the presence of a Doppler shadow, but this could be possibly due to a normalisation artifact \citep[e.g.,][]{2010Natur.465.1049S}. 
\citet{2020arXiv201107888Y} do not find any signs of pre-transit absorption in their H$\alpha$ analysis of WASP-33b, but we note that they looked in the planetary restframe only, while we noticed this PTS while looking in the stellar restframe.

Many photometric transit observations of WASP-33b have been gathered (see Sect.~\ref{sec:intro}), and even if they are affected by the stellar pulsations, none of them reported a pre-transit reduction of flux coherent with the planetary orbital period.
The most probable option in our opinion is that the PTS is still a stellar pulsation mode, which is possibly/probably excited by the planetary companion just before the transit. The PTS likely belongs to the same pulsations pattern well visible on H$\alpha$ and H$\beta$ lines in \citet[][see their Fig. 5]{cauley2020}, and also noticeable in Fig. \ref{fig:tomo_standard}, right panel. 
When averaging different transits the PTS does not average out like the other pulsations, but tends to clearly stand out beyond the noise. This would be indeed evidence of a somewhat expected star-planet interaction for this system.
Mechanisms that can excite stellar pulsations could have tidal or magnetic origin.

Planet-induced stellar pulsations were reported for example in the HAT-P-2 system, where \citet{2017ApJ...836L..17D} discovered pulsation modes corresponding to exact harmonics of the planet’s orbital frequency, indicative of a tidal origin.
Planetary induced tides in the host star, manifesting as the second harmonics of the orbital frequency, were proposed also for the systems WASP-12 and WASP-18 \citep{2020ApJ...889...54M,2020AcA....70....1M}.
Tidally induced flux modulations are also shown in {\it heartbeat} stars \citep[e.g.,][]{2011ApJS..197....4W,2012ApJ...753...86T}, caused by large hydrostatic adjustments due to the strong gravitational distortion they experience during the periastron passage of the eccentric binary companion. 
We thus can hypothesize a tidal star-planet interaction to be the cause of the PTS.
When considering equilibrium tides, we should find two RV maxima/minima during the planetary orbit, but the effect should be of the order of $\sim$10 \ms\ (i.e., lower than the one seen for WASP-18 considering the masses and orbital configuration of the two systems), thus not detectable.
When dealing with pulsations excited by tides instead, their amplitude could be much larger, but the lack of a stellar pulsation frequency close to the expected period of tides ($\sim$1.6 days, assuming a stellar rotational period of 0.884 days and a planetary orbital period of 1.22 days) in the frequencies shown in \citet{2020arXiv200410767V} is not in favour of this hypothesis.
Moreover, the duration of a resonance between pulsations and tides should be very short with respect to the evolutive timescales of the system. The large oscillations produced by the resonance would in fact dissipate a lot of energy, producing an exchange of angular momentum between rotational and orbital motion that modifies both, moving the system out of the resonance soon. So the probability that we are observing the system exactly in this moment is low, which is also not favouring the tidal hypothesis.

Another possibility for the detected PTS is that of a magnetic interaction. Magnetic star-planet interaction has been proposed and observed for some short-period systems, mainly as the modulation of stellar activity index with the orbital period \citep[e.g.,][]{2009A&A...505..339L,2012A&A...544A..23L,2018haex.bookE..25S,2019NatAs...3.1128C}.
A pre-transit absorption in the Balmer lines was observed for HD 189733b \citep{2015ApJ...810...13C} and was proposed as being caused by the planetary magnetic field, with a bow shock forming around the magnetosphere heading ahead of the planet \citep{2013MNRAS.436.2179L,2015ApJ...810...13C}.
Magnetic fields are known to have effect on stellar pulsations, depending on their magnitude \citep[e.g.,][and references therein]{2013pss4.book..207H}.
The interesting possibility of the PTS being a magnetic excitation of pulsations given by the interaction with the planetary magnetic field could be further explored with spectro-polarimetric measurements, since magnetic fields affect not only the spectral line profiles but also polarization properties of stellar radiation \citep[e.g.,][]{landi2004,2020ApJ...890...88O}.

%______________________________________________________________
\section{Cross-correlation with templates\label{sec:crosscorr}}

Since we found the K$_{\rm PTS}$ to be almost double of the planetary K$_P$ (Sect. \ref{sec:pretransit}), we tried to test a possible relation of the PTS with the planetary atmosphere, which could have left aliases or reflections/scattering in the spectra after removing the stellar master. We thus looked in our planetary transmission spectrum for species that could be present only in the planet atmosphere and not in the star, to be further investigated in the PTS restframe in case of detection.
We decided to look for Vanadium (\ion{V}{i}), which has been already detected in an exoplanetary atmosphere and not in the host star \citep{2020arXiv200605995B,borsaespresso}, and for Aluminum oxide (AlO), which has been claimed to be present in the atmosphere of WASP-33b through low-resolution spectrophotometric observations \citep{2019A&A...622A..71V}.

Planetary atmosphere transmission models were created by using \texttt{petitRADTRANS} \citep[pRT,][]{pRT}. 
We assumed solar abundances, a continuum pressure level of 1 mbar, and an atmospheric temperature-pressure profile calculated for the planet in the same way as presented in \citet{2020A&A...643A.131F}.
These parameters were set because they are typical for UHJs \citep[e.g.,][]{2019arXiv190502096H,stangret}. 
The atmospheric models, created in planetary radius as a function of wavelength, were translated into flux (R$_{\rm p}$/R$_{\rm s}$)$^2$, convolved at the HARPS-N resolving power and continuum normalized. 
While for \ion{V}{i} we used the opacities already included within pRT, for AlO we used those available at the opacity database of the Exoplanet Simulation Platform\footnote{https://dev.opacity.iterativ.ch/} \citep{2015ApJ...808..182G}, that we arranged in the pRT format.

Cross-correlation between the data and the models was performed in the stellar restframe on single residual spectra after the removal of a master out-of-transit star and telluric contamination (with the procedures explained in Sect. \ref{sec:Halpha_main}) on the whole wavelength range. This was done by working on the bi-dimensional $e2ds$ HARPS-N spectra order-by-order.
We define the cross-correlation as

\begin{equation}
C(v,t)= \sum\limits_{i=1}^N x_i(t,v) T_i
\label{equazccf}
\end{equation}
where $T$ is the model template normalized to unity and $x$ are the normalized fluxes at the $N$ wavelengths $i$ of the spectra taken at time $t$ and shifted at velocity $v$.
In this way we preserve the flux information \citep[e.g.,][]{2019arXiv190502096H}. 
In our models we impose all the values with contrast less than 5\% of the maximum in the wavelength range to be zero.

We selected a step of 1 \kms\ and a velocity range [-300,300] \kms,
performing the cross-correlation for each order. 
We performed a 5$\sigma$-clipping and masked the wavelength ranges most affected by telluric contamination. Then for each exposure we applied a weighted average between the cross-correlations of the single orders, 
where the weights applied to each order are the inverse of the standard deviation (i.e., the larger the S/N, the higher the weight) times the depths of the lines in the model template. 
For a range of K$_{\rm p}$ values from 0 to 300 \kms (with steps of 1 \kms) we averaged the in-transit cross-correlation functions after shifting them in the planetary restframe. This last step is done by subtracting the planetary radial velocity calculated for each spectrum as $v_{p}=K_{\rm p} \times \sin{2\pi \phi}$, with $\phi$ the orbital phase. 
In this way, we created the K$_{\rm p}$ vs V$_{\rm sys}$ maps, that are used to test the planetary origin of any possible signal.
We evaluated the noise by calculating the standard deviation of the K$_{\rm p}$ vs V$_{\rm sys}$ maps far from where any stellar or planetary signal is expected, in particular where $|V|>90$ \kms.

We find no evidence supporting the presence in the planetary atmosphere either of \ion{V}{i} or AlO, giving upper limits (which are model dependent) of 97 ppm and 17 ppm at the 3$\sigma$ level, respectively. 
While for \ion{V}{i} we are confident in the accuracy of the line list we used, as it has also already brought a clear detection \citep{borsaespresso}, this is not the case for AlO. 
We performed injection of our AlO model in our data using the abundance found in the detection by \citet{2019A&A...622A..71V}, and we could recover it with $\sim$14$\sigma$ significance. 
Although we did not find evidence for its presence with the available opacities, once an accurate 
AlO linelist will be available verifying with high-resolution spectroscopy the claimed presence of AlO in the planetary atmosphere \citep{2019A&A...622A..71V} will be possible using this dataset.

%______________________________________________________________
\section{Summary and conclusions\label{sec:discussion}}

We provide further evidence that WASP-33 is undergoing nodal precession, finding a precession period of $1108 \pm 19$~yr, slightly longer than that determined by \citet{Johnsonetal15}.
We also found that the stellar spin is perpendicular to the line of sight, and determined the gravitational quadrupole moment of the star $J_{2} = (6.73 \pm 0.22) \times 10^{-5}$, which is in close agreement with the theoretically predicted value of $(6.53 \pm 0.41) \times 10^{-5}$ \citep[e.g.,][]{RagozzineWolf09}.
By increasing the time coverage of WASP-33b spectroscopic transit observations it will be possible to constrain even more the precession period and possibly, with the growing precision of new-generation instruments, also to detect relativistic effects \citep{2016MNRAS.455..207I}. It would be important to look for other exoplanets in which we can detect orbital precession (other than WASP-33b and Kepler-13Ab), to increase the statistics and understand if this phenomenon could depend also on the presence of other companions, on the value of the stellar rotation or on the orbital obliquity of the system.

We found that stellar pulsations contaminate the extracted transmission spectrum of the planet and proposed a method to detrend it, which can be potentially applied to mitigate also stellar activity.
The detected H$\alpha$ absorption in the planetary atmosphere extends up to $\sim$1.18 R$_{\rm p}$, while absorption in 
\ion{Ca}{ii} H\&K was detected with an effective radius of $\sim$1.56 R$_{\rm p}$ \citep{2019A&A...632A..69Y}. This confirms the tendency of \ion{Ca}{ii} H\&K and H$\alpha$ to exist up to the highest layers of the atmospheres of hot Jupiters. 
The line contrast of the planetary absorption we measured for both H$\alpha$ and H$\beta$ lines is lower than previously found in the literature \citep{2020arXiv201107888Y,cauley2020}, while their contrast ratio is the same. This opens the possibility that the atmosphere of WASP-33b could be sensitive to a variable level of activity of the star. 

We detected a pre-transit signal almost coherent with the orbital period of the planet. The most likely explanation is that this is a stellar pulsation mode, which is excited by the planetary companion. The nature of this possible star-planet interaction is still doubtful, even if we tend to exclude tidal interactions and are more in favour of possible magnetic interactions.
Spectro-polarimetric observations of the transit and a complete spectroscopic phase coverage, as well as a detailed analysis of stellar pulsations on the line profiles, could help in understanding the nature of this pre-transit signal.

%______________________________________________________________

\begin{acknowledgements}
We thank the referee for their useful comments that helped improving the clarity of the manuscript. 
We acknowledge the support by INAF/Frontiera through the "Progetti Premiali" funding scheme of the Italian Ministry of Education, University, and Research and from PRIN INAF 2019.
FB acknowledges financial support from INAF through the ASI-INAF contract 2015-019-R0.
\end{acknowledgements}

%______________________________________________________________

\begin{appendix}
%\appendix
\section{Angular momentum geometry and precession in the WASP-33 system}
\label{app_precession}

\subsection{Tidal timescales}
\label{tidal_timescales}
The tidal timescales of the decay of the obliquity of the WASP-33 system and the exchange of angular momentum between the stellar spin and the orbit can be estimated using the tidal model of \citet{Leconteetal10} that we modify to make use of the stellar and planetary modified tidal quality factors $Q^{\prime}_{\rm s}$ and $Q^{\prime}_{\rm p}$, respectively. We adopt $Q^{\prime}_{\rm p}= 10^{5}$, a typical  value for a Jupiter-like planet, and $Q^{\prime}_{\rm s} = 5\times 10^{6}$, that corresponds to a strong tidal coupling as expected in late-type stars, but certainly underestimates the tidal timescales for an A-type star \citep{Ogilvie14}. Possible resonances between tides and stellar pulsations, that could increase the tidal dissipation inside the star, have been excluded by \citet{Kovacsetal13}. 

We find a timescale for the decay of the orbital eccentricity of only 0.2~Myr owing to the strong tidal dissipation enforced by the massive star inside the planet, while the decay of the obliquity and the spin-orbit exchange occur on timescales of $\sim 300$~Myr and $\sim 1$~Gyr, respectively, because they depend on the much smaller tidal dissipation inside the star. In conclusion, the precession in WASP-33 can be modelled by assuming that the obliquity and the stellar and orbital angular momenta are of constant magnitude, while the orbit can be assumed to have been already circularized \citep[cf.][]{DamianiLanza11}. 

\subsection{Determining the geometry of the system}
\label{determining_system_geometry}
The obliquity $\epsilon$ of the system is the angle between the angular momentum of the orbital motion and the spin of the star and is given by \citep[e.g.,][]{DamianiLanza11,Johnsonetal15}:
\begin{equation}
\cos \epsilon = \cos i \cos i_{\rm s} + \sin i \sin i_{\rm s} \cos \lambda,
\label{eqa1}
\end{equation}
where $i$ is the inclination of the orbital plane, that is, the angle between the line of sight $\hat{\bf k}$ (i.e., the direction from the observer on the Earth to the barycentre of the system) and the orbital angular momentum ${\bf l}_{\rm orb}$, $i_{\rm s}$ the inclination of the stellar spin axis to the line of sight, and $\lambda$ the spin-orbit angle projected onto the plane of the sky, that is measured by fitting the Rossiter-McLaughlin effect or modelling the motion of the Doppler shadow of the planet during its transits. The obliquity $\epsilon$ can be assumed constant on the timescales characteristic of the precession  (cf. Sect.~\ref{tidal_timescales}). 

\begin{figure}
\hspace*{-7mm}
\centering{
\includegraphics[width=8cm,height=7cm,angle=0]{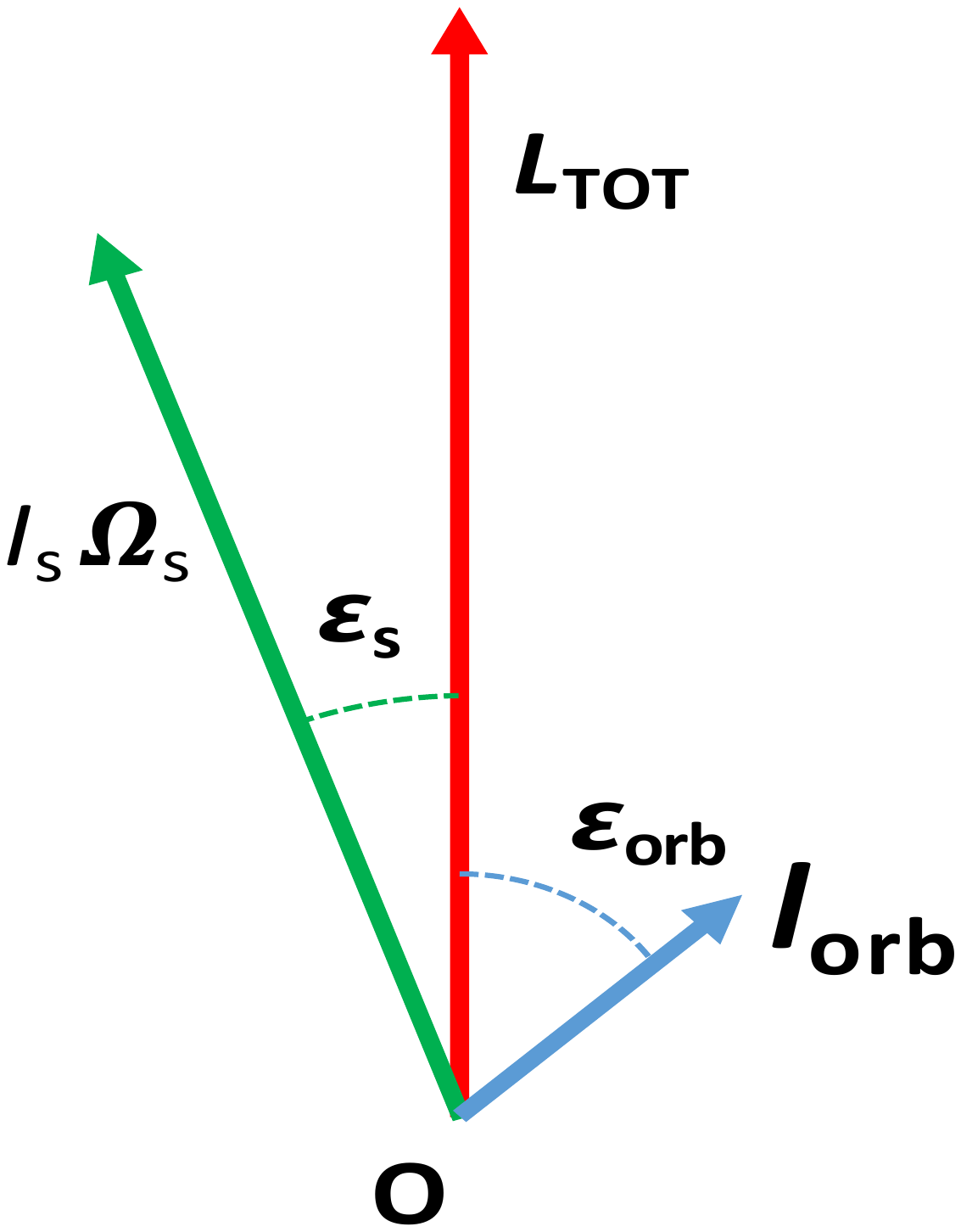}} 
\vspace*{-10mm}
   \caption{The total angular momentum of the system ${\bf L}_{\rm tot}$ (in red) as resulting from the composition of the stellar spin angular momentum $I_{\rm s} {\bf \Omega}_{\rm s}$ (in green) and the orbital angular momentum ${\bf l}_{\rm orb}$ (in blue). The angles $\epsilon_{\rm s}$ and $\epsilon_{\rm orb}$ between the stellar spin or the orbital angular momentum and the total angular momentum, respectively, are indicated. The obliquity of the system is $\epsilon = \epsilon_{\rm s} + \epsilon_{\rm orb}$, while its projection on the plane of the sky is $\lambda$ (not indicated).}
              \label{ang_mom_f1}%
\end{figure}

The total angular momentum of the system ${\bf L}_{\rm tot}$ is given by the sum of the stellar spin and the orbital angular momentum and  is conserved in the absence of external forces (see Fig.~\ref{ang_mom_f1}). It can be written as:
\begin{equation}
{\bf L}_{\rm tot} = I_{\rm s} \hat{\bf \Omega}_{\rm s} + {\bf l}_{\rm orb}, 
\label{eqa2}
\end{equation}
where $I_{\rm s} = M_{\rm s} (\gamma R_{\rm s})^{2}$ is the moment of inertia of the star of mass $M_{\rm s}$, equatorial radius $R_{\rm s}$, and gyration radius $\gamma$, while ${\bf \Omega}_{\rm s}$ is the stellar spin angular velocity that is related to the spectral line rotational broadening as $v \sin i_{\rm s} = \Omega_{\rm s} R_{\rm s} \sin i_{\rm s}$.  
%%%%%%%%%%%%%%%%%%%%%%%%%%%%%%%%%%%%%

The modulus of the total angular momentum can be found by taking the scalar product of eq.~(\ref{eqa2}) by itself and considering that $\cos \epsilon = \hat{\bf \Omega}_{\rm s} \cdot \hat{\bf l}_{\rm orb}$:
\begin{equation}
L_{\rm tot}^{2} = I_{\rm s}^{2} \Omega_{\rm s}^{2} + 2 I_{\rm s} \Omega_{\rm s} l_{\rm orb} \cos \epsilon + l_{\rm orb}^{2},
\label{eqltot}
\end{equation}
where the modulus of the orbital angular momentum for a circular orbit is:
\begin{equation}
l_{\rm orb} = m n a^{2}
\label{eqlorb}
\end{equation}
where $m = M_{\rm s} M_{\rm p} /(M_{\rm s} + M_{\rm p})$ is the reduced mass of the system, $M_{\rm p}$ the mass of the planet, $a$ the semimajor axis of the orbit, and $n = 2\pi/P_{\rm orb}$ the orbital mean motion with $P_{\rm orb}$ being the orbital period. 

Since ${\bf L}_{\rm tot}$ is constant, taking the scalar product of eq.~(\ref{eqa2}) by the unit vector directed along the line of sight $\hat{\bf k}$ and making the derivative with respect to the time gives:
\begin{equation}
\frac{di_{\rm s}}{dt} = - \frac{l_{\rm orb} \sin i}{I_{\rm s} \Omega_{\rm s} \sin i_{\rm s}} \left( \frac{di}{dt} \right),  
\label{eqa3}
\end{equation}
where $\cos i = \hat{\bf k} \cdot \hat{\bf l}_{\rm orb}$ and $\cos i_{\rm s} = \hat{\bf k} \cdot \hat{\bf \Omega}_{\rm s}$. 

Taking the time derivative of eq.~(\ref{eqa1}), making use of eq.~(\ref{eqa3}), and considering the  system at the particular epoch when $d\lambda / dt = 0$ and $di/dt \not= 0$ as happened in 2011, we obtain an equation that can be solved for $i_{\rm s}$, the inclination of the stellar spin to the line of sight:
\begin{eqnarray}
 \frac{l_{\rm orb} \sin i}{I_{\rm s} \Omega_{\rm s} \sin i_{\rm s}} \left( \sin i \cos i_{\rm s} \cos \lambda - \cos i \sin i_{\rm s} \right) & +  & \nonumber \\
\sin i \cos i_{\rm s} - \cos i \sin i_{\rm s} \cos \lambda & = & 0 
\label{eqa4}
\end{eqnarray}
The solution of eq.~(\ref{eqa4}) requires a few iterations because $\Omega_{\rm s} $ is derived from the spectroscopic $v \sin i_{\rm s}$ assuming that the stellar radius $R_{\rm s}$ is known. The numerical solution converges very rapidly and we adopt the solution closer to $\pi/2$ in our model. In principle, the method can be extended to any  epoch when $d\lambda/dt$ and $di/dt$ can both be measured, but the advantage of considering an epoch when $d\lambda/dt = 0$ is that of reducing the errors because the equations simplify and only $di/dt \not=0 $ is required. 

\begin{figure}
\hspace*{-7mm}
\centering{
\includegraphics[width=8cm,height=7cm,angle=0]{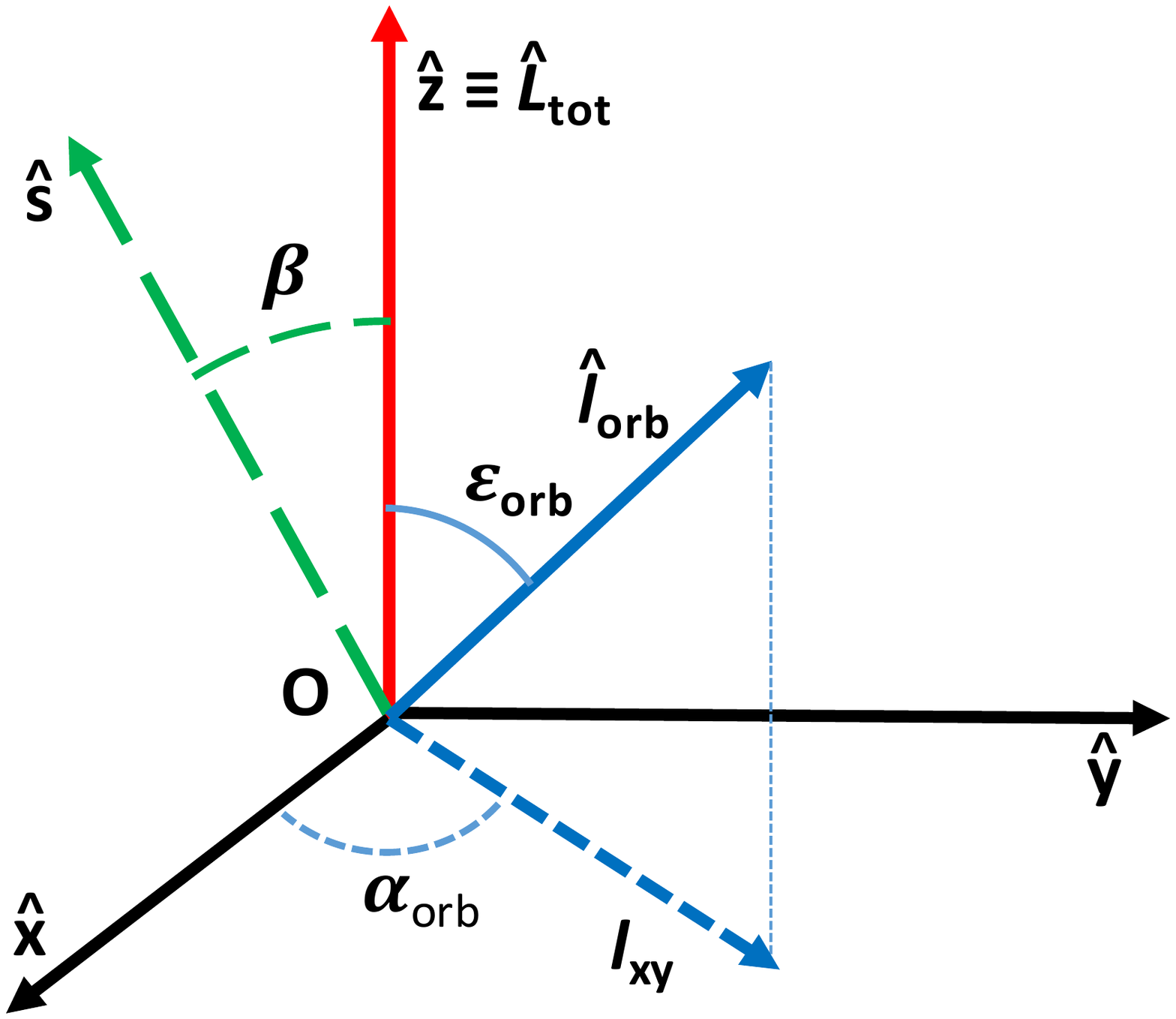}} 
\vspace*{-10mm}
   \caption{Reference frame adopted to specify the geometry of the WASP-33 system. The origin $O$ is at the barycentre of the system; the $\hat{\bf z}$ axis is directed along the total angular momentum of the system (in red); the $\hat{\bf x}$ axis is in the plane defined by the line of sight and the total angular momentum. The vector $\hat{\bf s}$ (dashed in green) is directed opposite to the line of sight, i.e., from the barycentre of the system towards the observer on the Earth. The unit vector in the direction of the orbital angular momentum $\hat{\bf l}_{\rm orb}$ is depicted in blue (solid line) together with its projection $I_{\rm xy}$ on the $xy$ plane (dashed line) and the angle $\alpha_{\rm orb}$ between that projection and the $\hat{\bf x}$ axis. The angle $\alpha_{\rm orb}$ varies uniformly in our inertial reference frame due to the precession of the orbital plane (see the text). For simplicity, the stellar spin vector and its projection on the $xy$ plane are not shown. }
              \label{ang_mom_f2}%
\end{figure}

To derive the other angles that specify the geometry of the system, we adopt a Cartesian orthogonal reference frame $(Oxyz)$ with the origin $O$ at the barycentre of the system, the polar axis $\hat{\bf z}$ directed along the total angular momentum, and the $\hat{\bf x}$ axis in the plane defined by the vectors ${\bf L}_{\rm tot}$ and the line of sight (cf. Fig.~\ref{ang_mom_f2}). 

We introduce the angle $\beta$ between the total angular momentum ${\bf L}_{\rm tot}$ and the unit vector $\hat{\bf s}$ directed opposite to the line of sight, i.e.,  from the barycentre of the system towards the observer, viz. $\hat{\bf s} \equiv - \hat{\bf k}$, and $\cos \beta  = \hat{\bf L}_{\rm tot} \cdot \hat{\bf s}$. We introduce also the angles $\epsilon_{\rm s}$ and $\epsilon_{\rm orb}$ between the total angular momentum and the stellar spin and orbital angular momentum, respectively, that is, $\cos \epsilon_{\rm s} = \hat{\bf L}_{\rm tot} \cdot \hat{\bf \Omega}_{\rm s}$ and $\cos \epsilon_{\rm orb} = \hat{\bf L}_{\rm tot} \cdot \hat{\bf l}_{\rm orb}$ (see Fig.~\ref{ang_mom_f1}). 
%%%%%%%%%%%%%%%%%%%%%%%%%%%%%%%%%%%%%
%%%%%%%%%%%%%%%%%
In our reference frame
\begin{equation}
\hat{\bf s} = (\sin \beta, 0 , \cos \beta). 
\label{eqs}
\end{equation}
Considering eq.~(\ref{eqa2}) and Fig.~\ref{ang_mom_f1}, we find:
\begin{equation}
\cos \beta  = (\hat{\bf s} \cdot {\bf L}_{\rm tot})/L_{\rm tot} = - (I_{\rm s} \Omega_{\rm s} \cos i_{\rm s} + l_{\rm orb} \cos i )/ L_{\rm tot}, 
\label{eqa5}
\end{equation}
\begin{equation}
\cos \epsilon_{\rm s} = ({\bf L}_{\rm tot} \cdot {\bf \Omega}_{\rm s}) /( L_{\rm tot} \Omega_{\rm s}) = (I_{\rm s} \Omega_{\rm s} + l_{\rm orb} \cos \epsilon) /L_{\rm tot},  
\label{eqa6}
\end{equation}
\begin{equation}
\cos \epsilon_{\rm orb} = ({\bf L}_{\rm tot} \cdot {\bf l}_{\rm orb})/( L_{\rm tot} l_{\rm orb}) = (I_{\rm s} \Omega_{\rm s} \cos \epsilon + l_{\rm orb}) / L_{\rm tot}. 
\label{eqa7}
\end{equation}
Equations~(\ref{eqa5}), (\ref{eqa6}), and (\ref{eqa7}) can be used to find the angles $\beta$, $\epsilon_{\rm s}$, and $\epsilon_{\rm orb}$ given that $\epsilon$ and $i_{\rm s}$ have been derived from equations~(\ref{eqa1}) and (\ref{eqa4}), $L_{\rm tot}$ comes from eq.~(\ref{eqltot}), and the inclination $i$ of the orbital plane is known from the model of the planetary transits.  

To complete the description of the  system, we need to derive the nodal angles $\alpha_{\rm s}$ and $\alpha_{\rm orb}$ between the $xz$ plane, defined by the vectors ${\bf L}_{\rm tot}$ and $\hat{\bf s}$,  and the spin and the orbital angular momentum, respectively. Since the stellar spin and the orbital angular momentum lie in the same plane containing ${\bf L}_{\rm tot}$, $\alpha_{\rm s} = \pi -\alpha_{\rm orb}$, so it is sufficient to derive the latter angle. The components of the unit vector directed along ${\bf l}_{\rm orb}$ can be expressed as (cf. Fig.~\ref{ang_mom_f2}):
\begin{equation}
\hat{\bf l}_{\rm orb} = (\sin \epsilon_{\rm orb} \cos \alpha_{\rm orb}, \sin \epsilon_{\rm orb} \sin \alpha_{\rm orb}, \cos \epsilon_{\rm orb}). 
\label{eqa8}
\end{equation}
Considering the definition of the cross product, we have:
\begin{equation}
(\hat{\bf l}_{\rm orb} \times \hat{\bf s})^{2} = \sin^{2} i. 
\label{eqa9}
\end{equation}
By expanding the l.h.s. of eq.~(\ref{eqa9}) by means of eqs.~(\ref{eqs}) and (\ref{eqa8}), we obtain an equation for $\alpha_{\rm orb}$:
\begin{equation}
A \cos^{2} \alpha_{\rm orb} + B \cos \alpha_{\rm orb} + C = 0,
\label{eqa10}
\end{equation}
where
\begin{eqnarray}
A & = & \sin^{2} \epsilon_{\rm orb} \sin^{2} \beta, \\ 
B & = & \frac{1}{2} \sin( 2\beta ) \sin (2 \epsilon_{\rm orb}), \\
C & = & \sin^{2} i - \cos^{2} \beta \sin^{2} \epsilon_{\rm orb} - \sin^{2} \beta. 
\label{eqa11}
\end{eqnarray}
Between the two solutions of eq.~(\ref{eqa10}), we choose  that giving the correct sign of 
\begin{equation}
\cos i = - \hat{\bf s} \cdot \hat{\bf l}_{\rm orb} = - \sin \beta \sin \epsilon_{\rm orb} \cos \alpha_{\rm orb} - \cos \beta \cos \epsilon_{\rm orb}. 
\label{eqa12}
\end{equation}
The values of the angles specifying the geometry of the WASP-33 system are listed in Table~\ref{table_w33_angles} together with their standard deviations as obtained by propagating the standard deviations of the system parameters and the systematic deviations obtained by shifting the epoch when $d\lambda/dt=0$ by 500 days after and before the assumed epoch, respectively.
\begin{table}
\caption{The angles and the factor $H(i, \beta, \epsilon, \alpha_{\rm orb})$ defining the geometry and the rate of nodal precession of the WASP-33 system at the epoch 19 October 2011 when $d\lambda/dt = 0$ is assumed. In addition to the standard deviations giving the statistical errors of the parameters, their systematic deviations obtained by shifting the epoch of $d\lambda/dt=0$ by 500 days after and before 19 October 2011 are listed, respectively.}
\label{table_w33_angles}
\begin{center}
\begin{tabular}{crcc}
\hline\hline
\noalign{\smallskip}
 & Value & Stand. dev.  & System. dev.\\ 
    & [degrees] & [degrees] & [degrees] \\
    \noalign{\smallskip}
\hline
\noalign{\smallskip}
 $i$ & 87.56 & 0.04 & \\
 $\lambda$ & -114.01 & 0.22  & \\
$\epsilon$ & 113.99 & 0.22 & $(-0.236; -0.669) $\\
$\epsilon_{\rm orb}$ & 92.37 & 2.73  & $(-0.239; -0.677)$\\
$\epsilon_{\rm s}$ & 21.63 & 2.72 & $(0.003; 0.008)$\\ 
$i_{\rm s}$ & 90.11 & 0.12 & $(-0.026; -0.015)$\\ 
$\beta$ & 90.84 & 0.17 & $(-0.135; 0.206)$\\
$\alpha_{\rm orb}$ & 92.47 & 0.06 & $(-0.414; 0.477)$ \\ 
$\alpha_{\rm s}$ & 87.53 & 0.06 & $(0.414; -0.477)$ \\
$H(i, \beta, \epsilon, \alpha_{\rm orb})$ &  -1.0021 & 0.0024 & $(2.1; 3.5) \times 10^{-4}$\\
\noalign{\smallskip}
\hline
\end{tabular}
\end{center}
\end{table}

\subsection{Precession of the orbital plane and the stellar spin}
\label{app_prec_formula}
The precession of the orbital plane is described by the precession of the vector ${\bf l}_{\rm orb}$ around the total angular momentum ${\bf L}_{\rm tot}$. In our reference frame, this precession corresponds to a uniform increase (or decrease) of the nodal angle $\alpha_{\rm orb}$ as a function of the time. We can compute the precession rate by differentiating eq.~(\ref{eqa12}) with respect to the time yielding:
\begin{equation}
\frac{d \alpha_{\rm orb}}{dt} = H(i, \beta, \epsilon, \alpha_{\rm orb})  \left( \frac{di}{dt} \right). 
\label{eqa13}
\end{equation}
where 
\begin{equation}
H(i, \beta, \epsilon, \alpha_{\rm orb}) \equiv -\frac{\sin i}{\sin \beta \sin \epsilon_{\rm orb} \sin \alpha_{\rm orb}}. 
\end{equation}
The factor $H(i, \beta, \epsilon, \alpha_{\rm orb})$ is very close to $-1$ in our system (cf. Table~\ref{table_w33_angles}), therefore the rate of nodal precession is virtually identical in absolute value to $di/dt$. 
The minimum and maximum values of the inclination of the orbital plane $i$, reached along a complete precession cycle, can be obtained from eq.~(\ref{eqa12}) for $\alpha_{\rm orb} = 0, \pi$ and are $\pi -(\beta \pm \epsilon_{\rm orb})$. 

Equation~(\ref{eqa13}) can be applied with the observed value of $di/dt$ to provide the observed rate of precession of the nodes of the planetary orbit; the precession period follows as $2\pi/(d\alpha_{\rm orb}/dt)$. Alternatively, it can be used with the theoretically derived $di/dt$  to predict the expected nodal precession rate in our system. Specifically, the theoretical expression of $di/dt$ for a circular orbit according to eqs.~(7) and~(11) of \citet{DamianiLanza11} is:
\begin{equation}
\frac{di}{dt} = {\cal A}_{\rm s} \Omega_{\rm}^{2} \cos \epsilon \sin i_{\rm s} \sin \lambda, 
\label{eq_didt}
\end{equation}
where
\begin{equation}
{\cal A}_{\rm s} \equiv \left( \frac{k_{2}}{n} \right) \left( 1 + \frac{M_{\rm p}}{M_{\rm s}} \right) \left( \frac{R_{\rm s}}{a} \right)^{5}, 
\end{equation}
where $k_{2}$  is the apsidal motion constant of the star that depends on its internal density stratification \citep[e.g.,][]{Claret19}.  
Note that the sign of $\lambda$ as defined by \citet{DamianiLanza11} is opposite to that adopted by \citet{Johnsonetal15} and by the present analysis;  eq.~(\ref{eq_didt}) has been reformulated according to the latter convention. 

The precession of the stellar spin occurs with the same rate as that of the orbital plane, but in the opposite direction as can be deduced from the conservation of the total angular momentum. The minimum and maximum inclinations of the stellar spin to the line of sight along a precession cycle can be derived as in the case of the orbital plane and are $\pi -(\beta \pm \epsilon_{\rm s})$. 

\subsection{Gravitational quadrupole moment of the star}
\label{app_quad_mom}
The precession rate can be related to the stellar gravitational quadrupole moment $J_{2}$ by means of eq.~(3) of \citet{Johnsonetal15} that we recast in our notation as:
\begin{equation}
J_{2} = - \frac{1}{3\pi} \frac{P_{\rm orb}}{\cos \epsilon} \left( \frac{a}{R_{\rm s}} \right)^{2}  \left( \frac{d \alpha_{\rm orb}}{dt} \right).  
\end{equation}
Note that our definition of the longitude of the ascending node of the orbit is different from that adopted by Johnson et al., but the precession rate is the same independently of the definition because the precession period is unique for the system. 

The gravitational quadrupole moment as derived from the observation of the precession of the orbital plane can be compared with the expectation from the theory.  Specifically, $J_{2}$ can be described as the sum of the two components $J_{2 \,\rm rot}$ and $J_{2\, \rm tide}$ produced by the centrifugal and the tidal distortion of the star, respectively. Their expressions are \citep[e.g.,][]{RagozzineWolf09}:
\begin{eqnarray}
J_{2} & = & J_{2 \, \rm rot} + J_{2 \, \rm tide},\\
J_{2 \, \rm rot}  & = & \frac{2k_{2}}{3} \left( \frac{\Omega_{\rm s}^{2} R_{\rm s}^{3}}{GM_{\rm s}} \right), \\
J_{2 \, \rm tide} & =  & k_{2} \left( \frac{M_{\rm p}}{M_{\rm s}} \right) \left( \frac{R_{\rm s}}{a} \right)^{3},
\label{j2_tide_exp}
\end{eqnarray}
where $G$ is the gravitation constant. In the case of WASP-33, $J_{2\, \rm tide} \sim 10^{-3} J_{2\, \rm rot}$, therefore it can be neglected in our analysis. Rigorously speaking, eq.~(\ref{j2_tide_exp}) is valid when the planet is on the equatorial plane of the star, but, given the smallness of $J_{2 \, \rm tide}$, this conclusion is not affected. 

Finally, we note that  we have considered the planetary spin synchronized and perpendicular to the orbital plane because tides can enforce such a configuration on a timescale as short as $\sim 1$~Myr \citep{Leconteetal10}. However, if this were not the case, we should include the contribution of the planetary gravitational quadrupole moment to the precession of the orbit as discussed by, e.g., \citet{RagozzineWolf09} and \citet{DamianiLanza11}. General Relativity precession is negligible in our case being $\sim 500$ times smaller than the nodal precession induced by the oblateness of the star \citep{Iorio11}. 

\subsection{A comparison with the model of \citet{2016MNRAS.455..207I}}
\label{appendix_iorio_model}

Another model of the precession in WASP-33 was presented by \citet{2016MNRAS.455..207I}. It assumes a Cartesian reference frame with the $y$ axis directed towards the observer, the $z$ axis along the projection of the stellar spin on the plane of the sky, and the $x$ axis perpendicular to both the other axes to form a right-handed orthogonal coordinate system \citep[see Fig.~1 of][]{2016MNRAS.455..207I}. In such a reference frame, the precession is described through the variation of the angles $\Omega$, the longitude of the ascending node of the orbit measured from the $x$ axis in the $x$-$y$ plane, and $I$, the angle between the $z$ axis and the orbital angular momentum. Note that the $x$-$y$ plane does not coincide with the plane of the sky, while the $x$-$z$ plane coincides with it. 

\begin{figure}[!ht]
\centerline{
\includegraphics[width=8.5cm,angle=90]{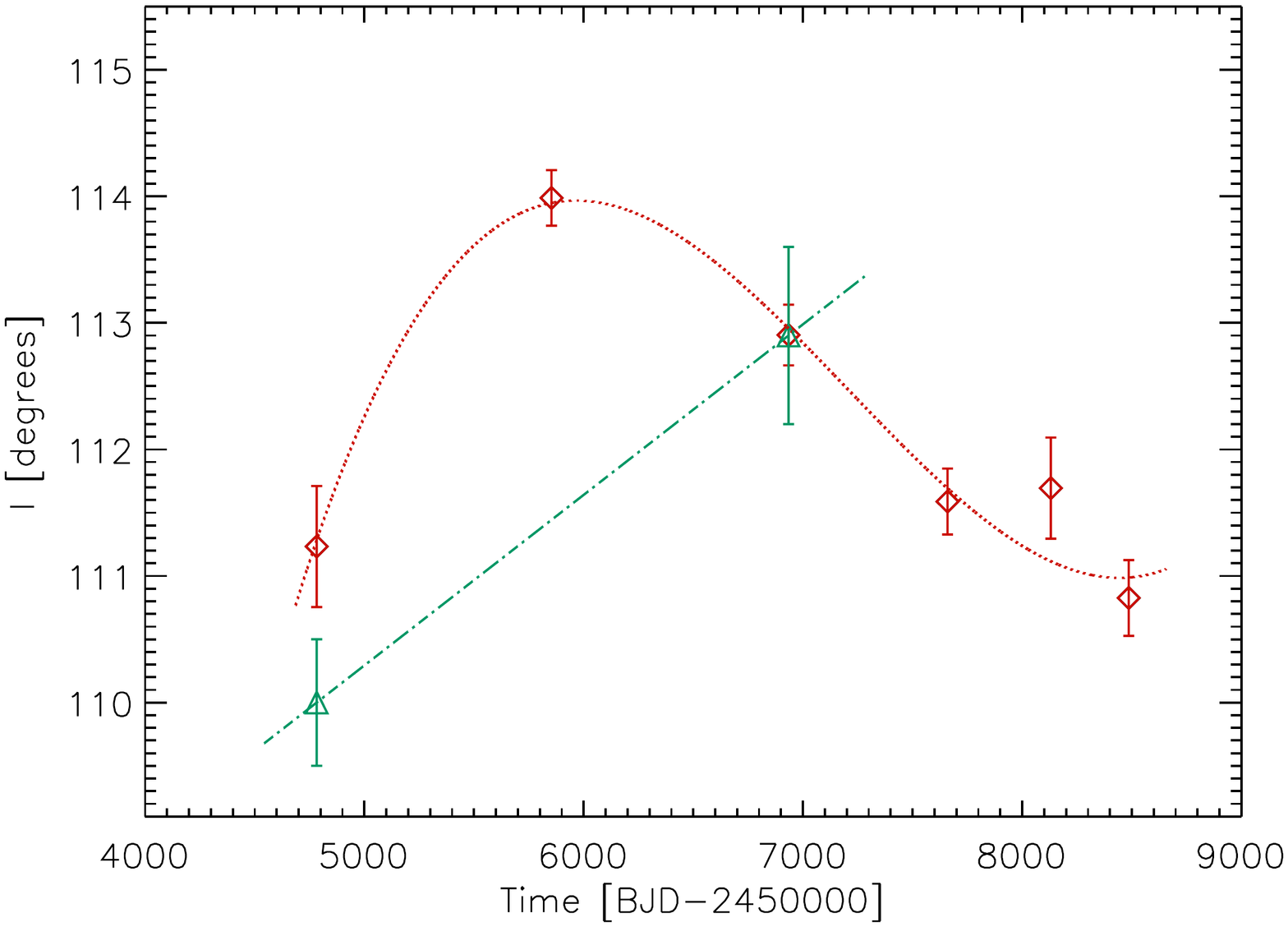}}
\caption{The angle $I$ of the model by \citet{2016MNRAS.455..207I} vs. the time. The red diamonds are the values found from our dataset, while the green triangles are the two values derived by Iorio from the data available to him at the time of his work. We see that the precession rate $dI/dt$ is not constant as shown by the cubic interpolation plotted as a dotted red line. For the sake of comparison, we plot a linear interpolation between the two values available to Iorio assuming a constant precession rate. }
\label{iiorio_vs_time}
\end{figure}

The two angles $\Omega$ and $I$ can be expressed in terms of the inclination $i$ of the orbital plane to the plane of the sky and of the projected obliquity $\lambda$ on the plane of the sky following equations from (25) to (33) of \citet{2016MNRAS.455..207I} as:
\begin{equation}
\Omega = \arccos \hat{N}_{x} \mbox{ for $\hat{N}_{y} \geq 0$ or}
\end{equation}
\begin{equation}
\Omega = 2\pi - \arccos \hat{N}_{x} \mbox{ for $\hat{N}_{y} <0$,}    
\end{equation}
where
\begin{equation}
\hat{N}_{x} =- \frac{\cos i}{\sqrt{\cos^{2} i + \sin^{2} i \sin^{2} \lambda}}    
\end{equation}
\begin{equation}
 \hat{N}_{y} = \frac{\sin i \sin \lambda}{\sqrt{\cos^{2} i + \sin^{2} i \sin^{2} \lambda}},
\end{equation}
while
\begin{equation}
I = \arccos (\sin i \cos \lambda).    
\label{eq_iiorio}
\end{equation}
Using the values of $i$ and $\lambda$ listed by \citet{Watanabeetal20} and those derived in the present work, we calculated the angles $\Omega$ and $I$ and listed them in Table~\ref{table_omega_i} together with their standard deviations obtained by propagating the errors on the inclination and the projected obliquity. 
\begin{table}[]
    \centering
    \begin{tabular}{rrrcc}
\hline\hline
\noalign{\smallskip}
Day & Month & Year & $\Omega$ & $I$ \\
 & & & [degrees] & [degrees] \\
 \noalign{\smallskip}
\hline
\noalign{\smallskip}
12  &    11  &  2008  &  274.00 $\pm$  0.078  &     111.23 $\pm$ 0.48 \\
19 &     10  &  2011  &  272.67 $\pm$  0.041   &    113.99 $\pm$ 0.22 \\
 4   &   10  &  2014  &  271.44 $\pm$  0.039  &     112.90 $\pm$ 0.24 \\
28  &     9  &  2016 &   271.11 $\pm$  0.043 &      111.59 $\pm$ 0.26 \\
12  &     1 &   2018  &  270.65 $\pm$  0.059   &    111.69 $\pm$ 0.40 \\
 2  &     1  &  2019 &   270.12 $\pm$  0.039 &      110.83 $\pm$ 0.30 \\
 \noalign{\smallskip}
\hline
    \end{tabular}
    \caption{The angles $\Omega$ and $I$ of the precession model of \citet{2016MNRAS.455..207I} on the epochs of our transit observations.}
    \label{table_omega_i}
\end{table}
The only two observations available to \citet{2016MNRAS.455..207I} were those of 2008 and 2014 with the parameters $i$ and $\lambda$ as reported by \citet{Johnsonetal15}, while we adopted the more precise determination of \citet{Watanabeetal20}. Therefore, Iorio estimated the rate of precession assuming a constant rate of change for both $\Omega$ and $I$ between those two epochs which led to an inclination of the stellar spin axis to the line of sight and a stellar quadrupole moment of, in our notation, $i_{\rm s} = 142^{\circ} \pm 11^{\circ}$ and $J_{2} = (2.1 \pm 0.8) \times 10^{-4} $, respectively. Nevertheless, as can be seen in Fig.~\ref{iiorio_vs_time}, the situation was different: the angle $I$ reached a maximum around 2011 and the precession rate $dI/dt$ became zero at that epoch. By differentiating eq.~(\ref{eq_iiorio}), we see that $dI/dt$ is zero close to the epoch when $d\lambda/dt= 0$, provided that $i$ is close to $90^{\circ}$ as in the case of a transiting planet, although the two epochs coincide only if $i=90^{\circ}$. 

Considering eq.~(9) of \citet{2016MNRAS.455..207I} and that his unit vector $\vec S^{*} =(0, \cos i_{\rm s}, \sin i_{\rm s})$ in our notation, we have for a circular orbit (cf. Section~\ref{tidal_timescales})
\begin{equation}
\frac{dI}{dt} = - \frac{3 n J_{2} R_{\rm s}^{2}}{2 a^{2}} \cos i_{\rm s} \sin \Omega \left( \sin i_{\rm s} \cos I - \cos i_{\rm s} \cos \Omega \sin I \right),
\label{diioriodt}
\end{equation}
where the symbols have been introduced above. Equation~(\ref{diioriodt}) for the given values of the angles $\Omega$ and $I$, implies that $dI/dt=0$ when $\cos i_{\rm s}= 0$ which confirms our result that the stellar spin is perpendicular to the line of sight around 2011. 

The precession rate of the angle $\Omega$ at the epoch when $dI/dt=0$ is given by eq.~(10) of \citet{2016MNRAS.455..207I}, that is
\begin{equation}
\frac{d\Omega}{dt} = \frac{3 n J_{2} R_{\rm s}^{2}}{2 a^{2}} \cos I
\label{domegadt}
\end{equation}
for a circular orbit. Considering our value of $J_{2} = (6.73 \pm 0.22) \times 10^{-5}$, we obtain a precession rate $d\Omega/dt = 0.325 \pm 0.015$ deg/yr in 2011 that is comparable with the observed mean precession rate $0.353 \pm 0.007$ deg/yr obtained by a linear regression over the six  epoches in Table~\ref{table_omega_i}. We do not expect a closer agreement because the precession rate $d\Omega/dt$ is not constant and the angle $I$ is also variable and affected by relatively large errors. Therefore, we prefer to estimate $J_{2}$ using equation~(3) of \citet{Johnsonetal15} because in that formula the precession rate and the true obliquity  $\psi$ are  constant, thus allowing a more precise determination. 

In conclusion, the differences between the results of \citet{2016MNRAS.455..207I} and ours as well as those of \citet{Watanabeetal20} are due to the paucity of observations available at that time leading to the assumption of constant precession rates for the angles $\Omega$ and $I$. Our more extended dataset allows us to show that the precession rate $dI/dt$ vanished around 2011, thus reconciling our results with those of Iorio's study and confirming that the inclination of the stellar spin axis to the line of sight was very close to $90^{\circ}$ at that epoch. 

\section{Radial velocities table\label{app_rv}}

\longtab[1]{
\renewcommand{\arraystretch}{0.8}
\begin{longtable}{ccc}
\caption{\label{tab:app_rv} HARPS-N RV observations of WASP-33.}\\
 \hline
 \hline
\noalign{\smallskip}
Time [BJD-2450000] & RV [\kms] & RV error [\kms]\\
\noalign{\smallskip}
\hline
\noalign{\smallskip}
\endfirsthead
\caption{Continued.} \\
\hline
\noalign{\smallskip}
Time [BJD-2450000] & RV [\kms] & RV error [\kms]\\
\noalign{\smallskip}
\hline
\noalign{\smallskip}
\endhead
\noalign{\smallskip}
\hline
\endfoot
\noalign{\smallskip}
\hline
\endlastfoot
7660.46966   &   -2.210  &   0.137 \\
7660.47683  & -2.181 & 0.148 \\
7660.48443 & -2.571 & 0.178 \\
 7660.49131 & -3.037 & 0.194\\
7660.49830  &   -3.026   &   0.224\\
7660.50580  &   -2.559   &   0.235\\
7660.51275  &   -1.997   &   0.221\\
7660.52008  &   -1.663   &   0.202\\
7660.52736  &   -2.144   &   0.222\\
7660.53481  &   -2.411   &   0.216\\
7660.54183  &   -2.756   &   0.172\\
7660.54914  &   -2.976   &   0.190\\
7660.55612  &   -3.029   &   0.173\\
7660.56343  &   -2.776   &   0.137\\
7660.57031  &   -2.720   &   0.132\\
7660.57806  &   -2.771   &   0.166\\
7660.58493  &   -2.583   &   0.187\\
7660.59189  &   -2.795   &   0.201\\
7660.60012  &   -3.039   &   0.176\\
7660.60686  &   -2.941   &   0.155\\
7660.61359  &   -2.244   &   0.173\\
7660.62139  &   -1.721   &   0.207\\
7660.62879  &   -1.975   &   0.177\\
7660.63575  &   -2.561   &   0.174\\
7660.64275  &   -2.914   &   0.175\\
7660.65028  &   -3.042   &   0.155\\
7660.65731  &   -2.902   &   0.156\\
7660.66485  &   -2.453   &   0.148\\
7660.67192  &   -2.933   &   0.143\\
7660.67904  &   -3.037   &   0.155\\
7660.68626  &   -3.153   &   0.156\\
7660.69377  &   -3.239   &   0.159\\
7660.70097  &   -3.226   &   0.150\\
7660.70797  &   -2.935   &   0.151\\
7660.71520  &   -2.797   &   0.160\\
7660.72256  &   -3.014   &   0.165\\
7660.72983  &   -3.380   &   0.168\\
7660.73690  &   -3.383   &   0.157\\
7660.74419  &   -3.137   &   0.158\\
7660.75355  &   -3.366   &   0.144\\
7682.42344  &   -2.566   &   0.111\\
7682.43060  &   -2.601   &   0.118\\
7682.43791  &   -2.730   &   0.129\\
7682.44515  &   -2.654   &   0.121\\
7682.45238  &   -2.555   &   0.122\\
7682.45945  &   -2.452   &   0.144\\
7682.46686  &   -2.718   &   0.164\\
7682.47396  &   -3.006   &   0.152\\
7682.48140  &   -2.915   &   0.160\\
7682.48844  &   -2.824   &   0.153\\
7682.49562  &   -2.810   &   0.147\\
7682.50304  &   -3.085   &   0.148\\
7682.51000  &   -2.913   &   0.170\\
7682.51738  &   -2.999   &   0.163\\
7682.52458  &   -2.988   &   0.177\\
7682.53171  &   -2.984   &   0.171\\
7682.53903  &   -2.963   &   0.167\\
7682.54636  &   -2.586   &   0.185\\
7682.55371  &   -2.800   &   0.220\\
7682.56072  &   -2.767   &   0.217\\
7682.56810  &   -2.708   &   0.208\\
7682.57520  &   -2.648   &   0.188\\
7682.58247  &   -2.614   &   0.199\\
7682.58974  &   -2.734   &   0.183\\
7682.59696  &   -2.590   &   0.164\\
7682.60426  &   -2.050   &   0.166\\
7682.66575  &   -3.309   &   0.191\\
7682.67303  &   -3.563   &   0.170\\
7682.68015  &   -3.158   &   0.159\\
7682.68730  &   -2.720   &   0.143\\
7682.69454  &   -2.764   &   0.128\\
7682.70348  &   -3.067   &   0.120\\
8131.32988  &   -2.929   &   0.138\\
8131.34056  &   -2.478   &   0.146\\
8131.35129  &   -2.453   &   0.138\\
8131.36176  &   -2.586   &   0.133\\
8131.37263  &   -2.864   &   0.120\\
8131.38317  &   -2.542   &   0.121\\
8131.39379  &   -2.386   &   0.124\\
8131.40434  &   -2.645   &   0.127\\
8131.41542  &   -3.192   &   0.142\\
8131.42616  &   -3.127   &   0.175\\
8131.43664  &   -2.762   &   0.171\\
8131.44760  &   -2.863   &   0.217\\
8131.45835  &   -2.928   &   0.186\\
8131.46897  &   -2.935   &   0.154\\
8131.47960  &   -2.682   &   0.164\\
8131.49031  &   -2.676   &   0.122\\
8131.50115  &   -2.434   &   0.127\\
8131.51169  &   -2.547   &   0.143\\
8131.52180  &   -3.232   &   0.140\\
8131.53303  &   -2.864   &   0.157\\
8131.54352  &   -2.565   &   0.165\\
8131.55425  &   -3.067   &   0.172\\
8131.56549  &   -3.110   &   0.163\\
8131.57598  &   -2.718   &   0.153\\
8131.58653  &   -2.855   &   0.143\\
8131.59718  &   -2.867   &   0.161\\
8486.31455  &   -2.277   &   0.144\\
8486.32180  &   -2.414   &   0.149\\
8486.32912  &   -2.613   &   0.143\\
8486.33632  &   -2.941   &   0.136\\
8486.34363  &   -3.336   &   0.135\\
8486.35089  &   -3.009   &   0.144\\
8486.35806  &   -2.761   &   0.159\\
8486.36541  &   -2.325   &   0.169\\
8486.37276  &   -2.603   &   0.188\\
8486.38002  &   -2.759   &   0.190\\
8486.38734  &   -2.627   &   0.183\\
8486.39448  &   -3.182   &   0.175\\
8486.40179  &   -3.122   &   0.193\\
8486.40892  &   -2.947   &   0.176\\
8486.41611  &   -2.984   &   0.159\\
8486.42364  &   -3.120   &   0.161\\
8486.43120  &   -2.873   &   0.158\\
8486.43789  &   -3.060   &   0.182\\
8486.44549  &   -3.294   &   0.160\\
8486.45276  &   -3.094   &   0.160\\
8486.46002  &   -3.065   &   0.198\\
8486.46729  &   -3.006   &   0.237\\
8486.47462  &   -2.418   &   0.200\\
8486.48194  &   -2.488   &   0.189\\
8486.48896  &   -2.484   &   0.168\\
8486.49661  &   -2.647   &   0.146\\
8486.50368  &   -2.734   &   0.147\\
8486.51103  &   -2.922   &   0.163\\
8486.51828  &   -2.520   &   0.150\\
8486.52578  &   -2.523   &   0.149\\
8486.53282  &   -2.601   &   0.133\\
8486.54008  &   -3.162   &   0.149\\
8486.54735  &   -3.237   &   0.161\\
\end{longtable}
}

\end{appendix}

\end{document}